\documentclass{aa}  

\usepackage{graphicx}
\usepackage{txfonts}
\usepackage{mathabx}
\usepackage{lipsum}
\usepackage{subcaption}
\usepackage{lscape}
\usepackage{placeins}
\usepackage{csquotes}
\usepackage{hyperref}
\usepackage{tabularx}
\usepackage{array}
\usepackage{soul}

\begin{document} 

   \title{Classical Cepheids in the Nuclear Stellar Disk: membership via
dynamical analysis}

   \author{
   M. De Leo \inst{1,2}
   \and
   M. Zoccali \inst{3}
   \and
   R. Albarrac\' in \inst{4,5}
   \and
   B. Acosta-Tripailao\inst{3}
   \and
   J. Minniti \inst{6}
    }

\institute{Dipartimento di Fisica e Astronomia, Università degli Studi di Bologna, Via Piero Gobetti 93/2, Bologna, 40129, Italy \\
        \email{micheledl89@gmail.com}
    \and
Osservatorio di Astrofisica e Scienza dello Spazio di Bologna, INAF, Via Piero Gobetti 93/3, Bologna, 40129, Italy
    \and
Instituto de Astrof\'isica, Pontificia Universidad Cat\'olica de Chile, Av. Vicu\~na Mackenna 4860, 782-0436 Macul, Santiago, Chile
    \and
Max Planck Institute for Astronomy, D-69117 Heidelberg, Germany
    \and
Fakultät für Physik und Astronomie, Universität Heidelberg, Im Neuenheimer Feld 226, 69120 Heidelberg, Germany
    \and
Department of Physics and Astronomy, Johns Hopkins University,
Baltimore, MD 21218, USA
}
   \date{Received XXX; accepted YYY}

 
  \abstract
   {The nuclear region of the Milky Way is a dust-obscured, observationally challenging and complex superposition of structures. Nevertheless, it can provide crucial information on the formation of the inner Galaxy, in particular for the star formation history as seen from the age distribution of the stars inhabiting the region. In this context, the precise age data gleaned from pulsating variables can be of fundamental importance.}
   {Four classical Cepheids have been associated to the Nuclear Stellar Disk (NSD), suggesting a burst of star formation $\sim$20 Myr ago. In this work, we analyse all available data for these four Cepheids in order to gauge if they are bona fide members of the NSD or interlopers in the region. This allows to confirm or rule out the star formation burst hypothesised in the literature.}
   {Our dynamical analysis is based on an updated MW potential that includes all components of the inner Galaxy. We conduct a thorough study of possible systematics, taking into account both observational errors and model dependencies.}
   {Our results suggest that only one Cepheid likely belongs to the NSD, while the orbits of other two extend well beyond it. Accounting for observational uncertainties, only $\sim$23–28\% of the realisations sourced from the error distributions for the latter two are compatible with the NSD. For the fourth there are no proper motion measurements available and the light curve analysis puts into question its classification as a classical Cepheid.}
   {Through a comprehensive dynamical analysis of available data on the Cepheids attributed to the NSD, we gauged their membership to this structure. Out of the three Cepheids analysed in unprecedented depth, GCC-a is likely a member of the NSD while both GCC-b and GCC-c do not seem dynamically linked to it. Factoring in the observational uncertainties, a non-negligible percentage of the realisations for GCC-b and GCC-c are still compatible with the NSD, thus membership cannot be completely excluded for either with current data. The robust association of only one Cepheid casts doubt on the occurrence of a star formation episode in the NSD, about 20 Myrs ago.}


   \maketitle

\section{Introduction} \label{intro}

The region of the Milky Way (MW) comprised within $\sim$3 kpc from the centre, generically called the "bulge", includes two main components: a bar and a spheroid. They obviously differ in shape \citep[e.g.,]{zoc+17_gibs3, lim+21} but also in the kinematics and chemical abundances \citep[e.g.,]{zoc+17_gibs3, queiroz+21}. Their age distribution is the subject of a long standing debate: while the metal poor spheroid is certainly as old as $\sim$10 Gyr, the metal rich bar may or may not include a significant fraction of stars with ages between 1 and 7 Gyrs \citep[see e.g.,][]{clarkson+18, Haywood+16, bensby+17, bernard+18, renzini+18}. The reader is referred to \citet{zoc24_chap} for a recent review of the bulge properties.

In contrast with this large, passive component the region within a radius of $\sim$200 pc in the plane, and a height of $\sim$45 pc perpendicular to it, includes the Central Molecular Zone (CMZ) with several active sites of star formation, a dense stellar component known as the Nuclear Stellar Disk (NSD), three massive clusters as young as $\sim$5 Myr, several powerful X-ray sources, radio bubbles, and the supermassive black hole \citep[see][for a recent review]{bryant+krabbe2021}.

The NSD, in particular, has this name because it is flattened and very dense in stars, resembling the NSDs of external spiral galaxies \citep{schultheis25}. It is natural, then, to expect that it should also be a fast rotating, cold component, with a stellar population younger and more metal-rich than the main bulge. Due to extreme interstellar extinction, however, the study of this dense and obscured stellar component has progressed very slowly, as it is only possible in the near infrared (IR), with exquisite spatial resolution and large collecting power. In addition, the color difference between main sequence stars and giants is smaller than the color spread induced by reddening variations, that are large even at scales of a few arcsecs \citep{nogueras-lara+21ext}. This is true for any filter combination, but especially dramatic in H$-$K, the two near IR filters that are most efficient, to map these highly obscured stars.

While the NSD certainly rotates \citep[e.g.,]{schonrich+2015}, what is not yet clear is whether it does so faster, and with lower dispersion, than the bulge just outside it \citep{zoc24_NSD}. Dedicated studies combining metallicity with kinematics show discrepant results, mainly due to the low statistics and the difficulty of isolating a pure NSD population from the many contaminants in front and behind it. While \citet{schultheis+21} found a metallicity dependent rotation velocity, suggesting that the NSD is fast rotating, and made up by metal-rich stars with [Fe/H]$\sim$$+0.3$, \citet{fk25} found no metallicity-dependent rotation velocity. Similarly, the sigma has been reported to be $\sigma_{\rm RV}=62 \,{\rm km \, s^{-1}}$ according to \citet{fk25}, or variable between $80 \,\, {\rm and} \,\, 120 \, {\rm km \, s^{-1}}$, for the super-solar or sub-solar metallicity stars, respectively \citep{schultheis+21}. For reference, the velocity dispersion just outside the NSD, along the bulge minor axis at b=$-1$ degree, is quite different, at about $130 \,{\rm km \, s^{-1}}$ \citep{valenti+18, quezada25}. In this respect, it is important to emphasise a fundamental difference between the mean rotation velocity and the observed velocity dispersion of a population of stars. The mean rotation velocity can be determined with good precision, even if individual measurements have large errors, provided we can measure the velocity of a statistically large sample of stars (as we do with proper motions, PMs). Meanwhile, the observed velocity dispersion is a convolution of the real dispersion with the observational error, often varying with magnitude, hence a good control of the errors is crucial to derive a reliable $\sigma_{\rm RV}$.

As for the age distribution, \citet{nogueras-lara+20} found that the NSD formed $\sim$90\% of its stars $>$8 Gyr ago, with the other $\sim$10\% having formed about 1 Gyr ago, or later (depending on the adopted stellar models). The presence of an age gradient was later suggested by \citet{nogueras-lara+23}, with the outer part of the NSD including a larger fraction of intermediate age stars. These results, however, are based on a comparison of the observed, dereddened luminosity function of the red giant branch (RGB) and red clump (RC) with a linear combination of stellar models. They are affected by large uncertainties, because RC stars are widely used standard candles, precisely due to the very weak dependence of their magnitude upon age and metallicity, especially in the near IR \citep[e.g.,][]{girardi+16}. 

Independent information about the age distribution of stars in the NSD comes from the study of variable stars, especially pulsating ones, as they obey a period-age relation. Therefore, they can be used as tracers of populations of different ages, if selected according to their periods. The large extinction complicates the analysis in this case, too: it affects both the classification of these variables and their distance determination. In addition, while proper motions are often derived from the same multi-epoch data used for the variability analysis, the derivation of radial velocities is expensive in terms of telescope time and it is available only for a limited sample. Information on the three dimensional (3D) velocity is essential to derive orbits, thus establish whether a given variable truly belongs to the NSD.

\citet{matsunaga+09miras} identified several Mira variables towards the NSD, and calculated distances for 143 of them, clumping at the distance of the Galactic centre (GC). About 20 of them have proper motions from Gaia DR3 \citep{gaiadr3} but they do not have measured radial velocities. Later, \citet{sanders+22} reported the discovery of $\sim$1800 Mira variables in the VVV survey, suggesting the presence of an important population of intermediate age stars, in the NSD. From the same data, however, \citet{albarracin+25} claimed that most of them are either not bona fide Miras, or they are located at the far side of the disk. It should be emphasised that, indeed, Miras that are truly located in the NSD are strongly saturated in VVV, but almost invisible in Gaia, further complicating the analysis. This is true, in particular, for all the Miras in the sample of \citet{matsunaga+09miras}.

The presence of classical Cepheids (with ages between $20$ and $200$ Myr), in the NSD would be highly valuable as an independent confirmation of the recent burst of star formation claimed by \citet{matsunaga+15} and \citet{nogueras-lara+20}. Indeed, \citet{matsunaga+11} identified three classical Cepheids in the NSD. \citet{matsunaga+15} added a fourth one, and measured radial velocities for all of them. According to these two studies, the four classical Cepheids identified are $\sim$$25$ Myr old. If all of them belonged to the NSD then it would be believable that they all originated in the same star formation episode, thus confirming a recent burst. The PMs catalogue from \citet{shahzamanian+22} includes measurements for three of these four Cepheids and while they do appear to fall within the NSD distribution, they also fall into the bulge and foreground distributions \citep[Figures 9, 11 and 12 from][]{shahzamanian+22}. We report in Appendix~\ref{PMs_RV} the distributions of the proper motion in Galactic longitude and of the Galactocentric velocity of both the NSD and inner bulge from the literature \citep{schultheis+21, shahzamanian+22, quezada25}, showing the substantial overlap of the kinematical signature of these Galactic components and how the Cepheid's values are compatible with both distributions. This underscores the need for a complete orbital modeling as individual phase-space measurements cannot properly discriminate between the overlapping structures present in the GC.

By putting together the original data \citep{matsunaga+11, matsunaga+13, matsunaga+15} with PMs \citep{shahzamanian+22} and an independent verification of the light curves from the VISTA Variables in the V\'\i a L\'actea (VVV) survey \citep{minniti+2010}, we performed a comprehensive analysis of the four classical Cepheids to confirm their status as members of the NSD.

The paper is organised as follows: in Sect.~\ref{data} we present the data used for our analysis and discuss their sources, in Sect.~\ref{method} we introduce the dynamical portion of the analysis carried out, of which we present the results in Sect.~\ref{results} while Sect.~\ref{concl} contains a discussion and our conclusions.

\section{Data} \label{data}

   \begin{figure}
   \centering
     \includegraphics[width=8.5cm]{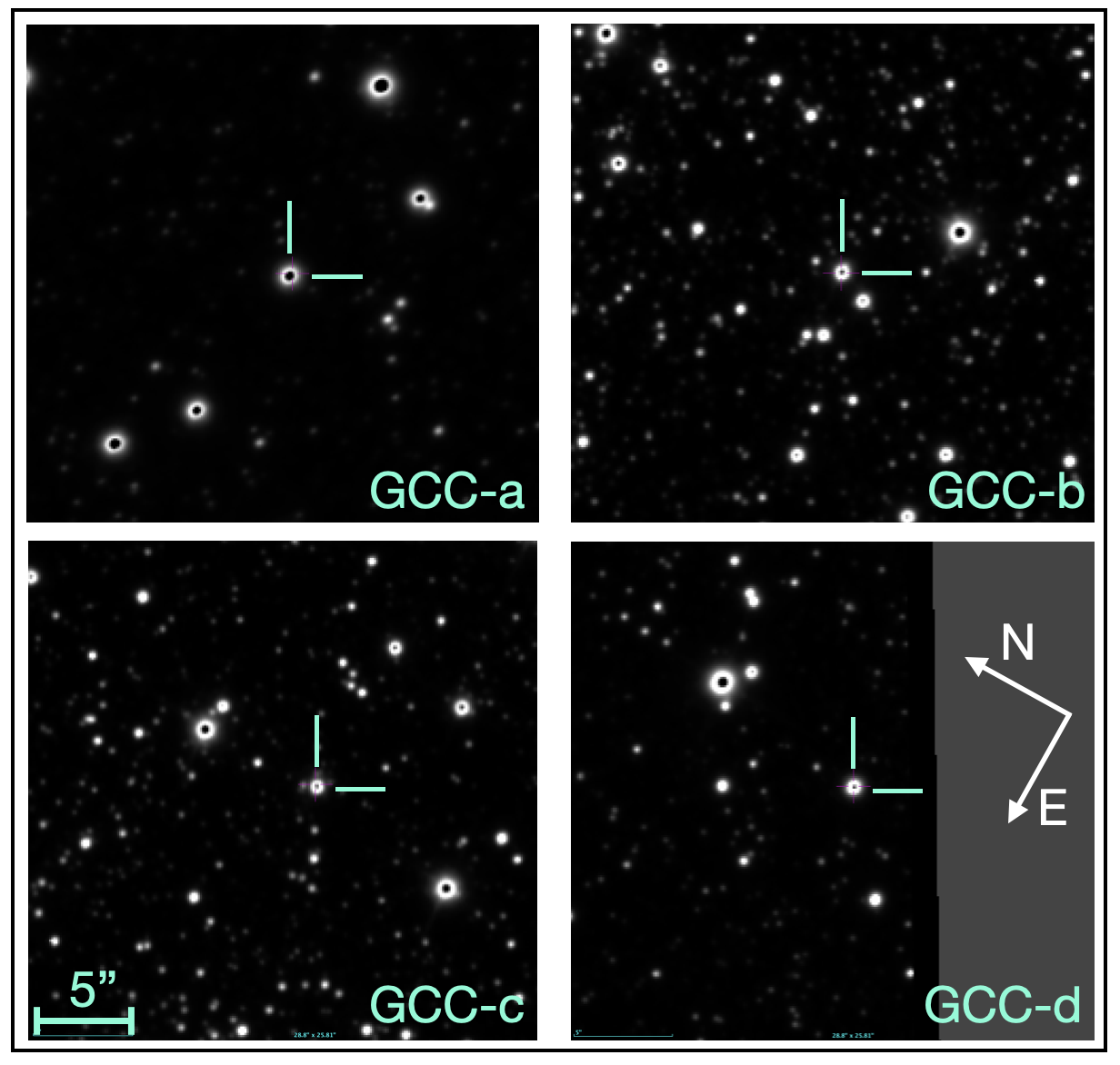}
      \caption{Finding charts. A $25^\second$$\times$$25^\second$ region is shown around each of the four Cepheids in \citet{matsunaga+15}. The images are from HAWK-I@VLT, in the K$_{\rm s}$-band. Star GCC-d falls at the edge of the frame, hence the grey region.}  
         \label{images}
   \end{figure}

We study the four classical Cepheids from \citet{matsunaga+15}. The latter provides the coordinates, H band light curves, and mean radial velocities for the four variables. Finding charts were published by \citet{matsunaga+11}, based on J, H and K false-color images taken with the SIRIUS camera at the Infrared Survey Facility 1.4-m telescope: the same used to obtain the light curves. The pixel scale was 0.45 arcsec/pix \citep{nagayama+03}, and the typical seeing between 1 and 1.3 arcsec. We update them here with K-band images from the HAWK-I camera at the Very Large Telescope (VLT) images, with a spatial resolution of 0.106 arcsec/pix and a FWHM of 0.3 arcsec, after correction by the Ground Layer Adaptive optics assisted by Lasers. Details of these observations are given in \citet{valenzuela+26}. These images allow us to verify that, even at the higher spatial resolution of HAWK-I (0.106 arcsec/pix), these stars are isolated, although saturated in these data.

\begin{figure}
   \centering
     \includegraphics[width=8.5cm]{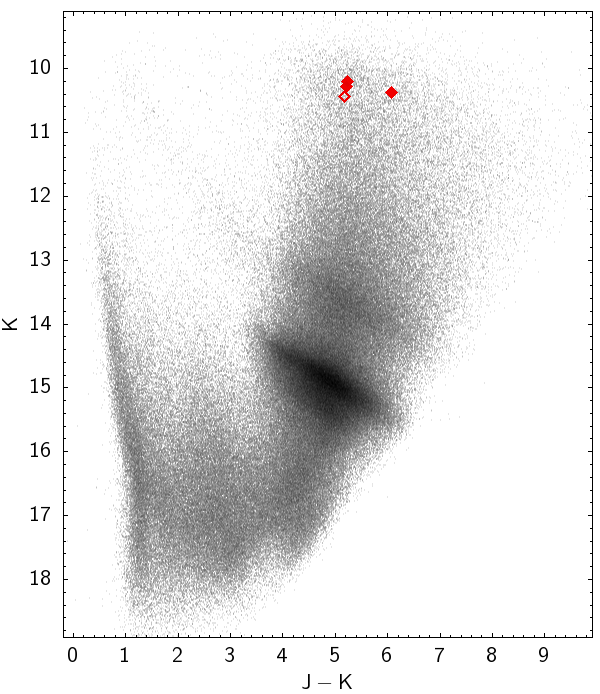}
      \caption{Color magnitude diagram. The magnitudes and colors of the four Cepheids are shown, in red, on top of the CMD from the entire "central" field, of the GALACTICNUCLEUS program (grey points). The open symbol refers to GCC-d, for which PM is not available.}
         \label{cmds}
   \end{figure}

All the variables are included in the "central" field of the GALACTICNUCLEUS program \citep{nogueras-lara+19}. A color magnitude diagram (CMD) of the whole "central" field is shown in Fig.~\ref{cmds}, where the Cepheids are marked with a red symbol. The diagram also shows the RC, as the main overdensity at K$\sim$15, spread diagonally along the reddening vector. The less prominent overdensity about 1 magnitude brighter is the Early Asymptotic Giant Branch, while the almost vertical structures at the blue end of the CMD are foreground disk main sequence stars. Although the intrinsic color of Cepheids variables is expected to be slightly bluer than that of the RC, the large reddening and its strong spatial variations are dominant, in this case, with respect to the intrinsic color. 

As previosuly stated, three of the four Cepheids are also present in the PM study by \citet{shahzamanian+22}, obtained by combining a first epoch from the Hubble Space Telescope Paschen-$\alpha$ survey, from 2008 \citep{wang+2010, dong+11}, and a second epoch, seven years later, from GALACTICNUCLEUS \citep{nogueras-lara+19}. These stars are marked as a filled symbol in the CMD, and they are the ones for which we will derive orbits hereafter. The fourth variable, GCC-d, is not included in the PM catalogue by \citet{shahzamanian+22}.

Finally, we use the distances reported in \citet{matsunaga+13}, assuming that the classification as classical Cepheids is correct. As Table 5 of \citet{matsunaga+13} shows, if the classification changes to Type II Cepheids, all three stars (numbers 18, 20 and 32) would be shifted at distances of $\sim2 \, {\rm kpc}$, incompatible with NSD membership.

We list in Table~\ref{tab_input} all the phase-space coordinates used to compute the orbits, with their accompanying uncertainties (discussed in Sec.~\ref{obserrs}).

\begin{table*}
 \caption{Input phase-space variables with uncertainties}\label{tab_input}
 \centering
  \begin{tabular}{c c c c c c c }
 \hline\hline
 ID & ${\rm l}$ & ${\rm b}$ & ${\rm Dist}$ & ${\rm RV}$ & $\mu_{\rm l}$ & $\mu_{\rm b}$   \\
  & $[^{\circ}]$ & $[^{\circ}]$ & ${\rm [kpc]}$ & ${\rm [km \, s^{-1}]}$ & ${\rm [mas \, yr^{-1}]}$ & ${\rm [mas \, yr^{-1}]}$  \\
 \hline \noalign{\smallskip}
 GCC-a &   +0.186 & $-$0.009 & $8.3\pm0.5$ &  +$127\pm13$ & $-3.81\pm0.92$ & $0.21\pm0.79$ \\
 GCC-b & $-$0.105 & $-$0.043 & $7.8\pm0.5$ & $-58\pm13$ & $-4.57\pm0.37$ & $1.77\pm0.38$ \\
 GCC-c & $-$0.112 & $-$0.041 & $7.7\pm0.5$ & $-81\pm13$ & $-0.43\pm0.58$ & $2.07\pm0.67$ \\
 \noalign{\smallskip}
 \hline
 \end{tabular}
 \tablebib{
 l, b and RV from \citet{matsunaga+15}, Dist from \citet{matsunaga+13}, $\mu_{\rm l} \, {\rm and} \, \mu_{\rm b}$ from \citet{shahzamanian+22}.
 }
\end{table*}

Concerning the photometric variability analysis, we were able to complement the light curves published by \citet{matsunaga+15} with data points from the VVV survey \citep{minniti+2010}. Point Spread Function photometry for VVV is publicly available from the VIRAC2 catalogue \citep{VIRAC2}. These stars, however, are very bright and saturated in many of the VVV images, hence not present in the VIRAC2 catalogue. Our group also carried on Point Spread Function photometry for the epochs between 2010 and 2019 of VVV, with the method described in \citet{contreras+17}. In our data, the magnitude of the Cepheids was correctly recovered in most of the epochs for all the bands available, K (reported here in Fig.~\ref{fig:Kband_LCs}) and J and H (in \ref{fig:LCs_JH} in Appendix ~\ref{multilc}) although the scatter is larger than in the data by \citet{matsunaga+15}. All our light curves show only the VVV data points with photometric errors lower than 0.1 magnitude. While VVV data do not improve the definition nor the sampling of the light curves, they do allow us to confirm the periods, especially given that they were taken several years later, hence a small error in the periods would become more visible after some years.

For GCC-d, however, the literature light curves were already poorly populated and noisier. The classification of this variable as a classical Cepheid was more uncertain than that of the other three, as shown in Fig.~\ref{fig:GCCd_LC}. Unfortunately, VVV data do not improve the analysis, either. If we add the VVV points to the literature ones, we recover the period of 18.886 days, quoted by \citet{matsunaga+15}. With VVV data alone, though, we would retrieve a slightly lower period (17.907 days). Given the poorer precision of the VVV data, we do not have strong arguments to favor such a shorter period. Nevertheless, independent confirmation of both its classification as a classical Cepheid and of its period, with additional photometric data would be desirable. In what follows, given that the PM for this variable is not present in the \citet{shahzamanian+22} catalogue, we will not analyse it further, as we would not be able to do so in an homogeneous manner compared to the other three. 

   \begin{figure}
   \centering
     \includegraphics[width=8.5cm]{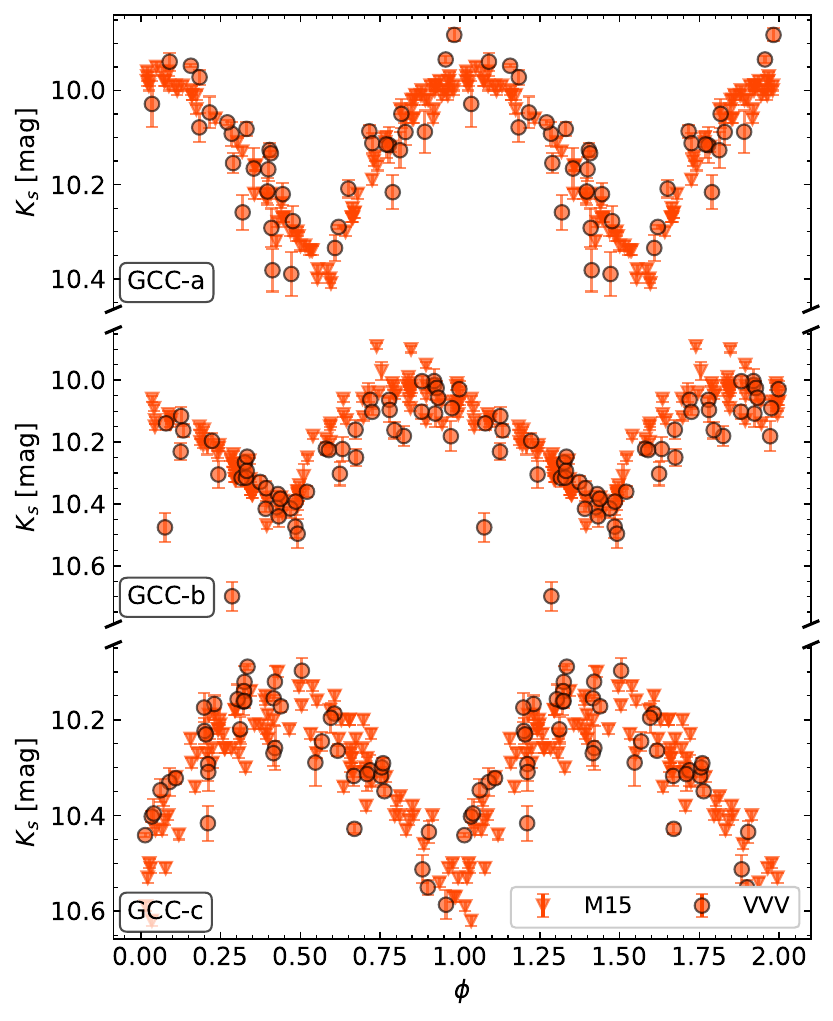}
      \caption{K band light curves for the three Cepheids for which we were able to calculate orbits. Triangles show the data from \citet{matsunaga+15}, while circles are data from our VVV photometry. }
         \label{fig:Kband_LCs}
         
   \end{figure}
   \begin{figure}
   \centering
     \includegraphics[width=8.5cm]{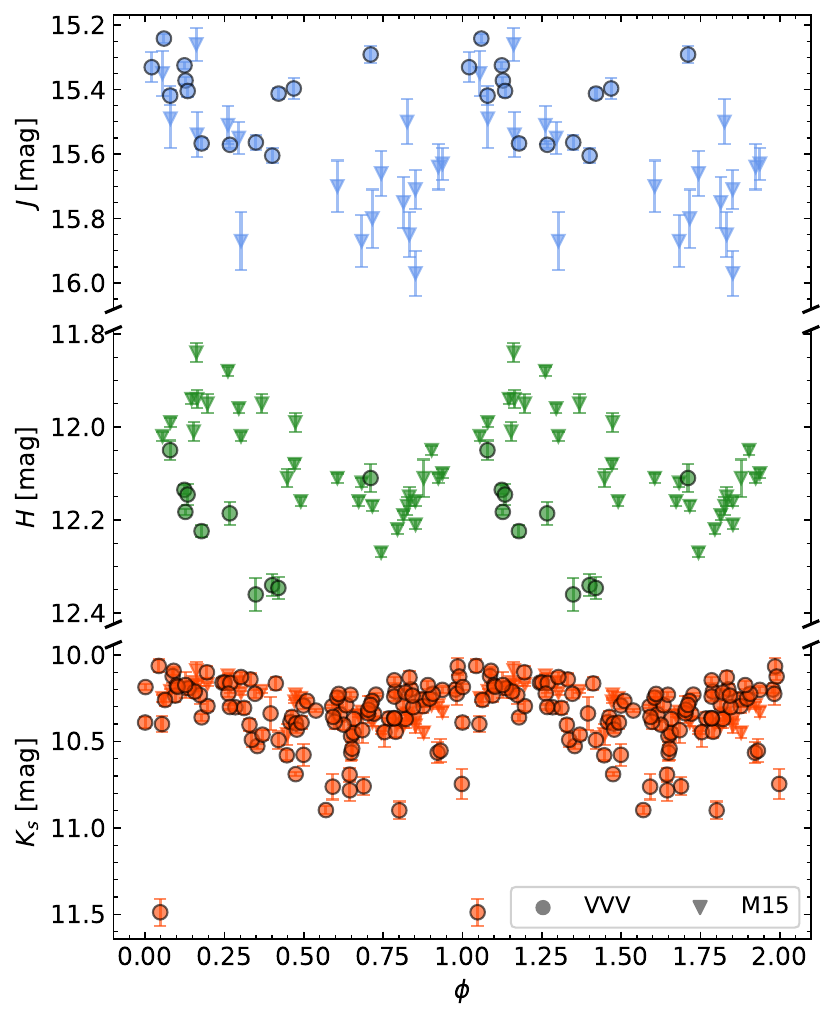}
      \caption{J, H and K band light curves for GCC-d. Triangles are from \citet{matsunaga+15}, while circles are from our VVV photometry }
         \label{fig:GCCd_LC}
   \end{figure}

\section{Method}\label{method}
Throughout this work, we carry out a full dynamical analysis of the three Cepheids mentioned in Sec.~\ref{data} with complete 6D phase-space information: GCC-a, -b, and -c. The tool that enabled to carry out this analysis is the in-house orbital integration code \textsc{OrbIT} \citep{julio+24,DeLeo+26_bar,DeLeo+26_clusters}.

\subsection{Characteristics of \textsc{OrbIT}}\label{orbit}
The code and its underlying potential model for the MW are fully presented in \citet{DeLeo+26_clusters}. Briefly, the MW is represented by a time-varying multi-component potential including: a Navarro-Frenk-White \citep[NFW, ][]{nfw1996} dark matter (DM) halo, two Miyamoto-Nagai \citep[MN, ][]{mndisc1975} discs for the stellar thin and thick discs, two MN discs for their gaseous counterparts, and a bulge composed of a Plummer \citep{plummer1911} spherical component and a Long-Murali \citep[LM, ][]{lmbar1992} rotating bar. The parameters of each component of this model are reported in \citep{DeLeo+26_bar,DeLeo+26_clusters}. To properly characterise the influence of the NSD on the dynamics of the Cepheids, we included it in the MW model as a MN disc with mass $M_{NSD}=1.05\cdot10^9 \, {\rm M_{\odot}}$, scale length $a_{NSD} = 88.6 \, {\rm pc}$ and scale height $b_{NSD} = 28.4 \, {\rm pc}$ following \citet{sormani+22}.

\textsc{OrbIT} is written in the computing language C using a \textquote{kick-drift-kick} leapfrog integration scheme \citep[i.e. a modified Verlet scheme, ][]{verlet1967,bt2008}, which is time-reversible and suppresses numerical errors \citep{rq1992,hlw2003}. Initial conditions for \textsc{OrbIT} were provided by transforming the observed positions, distances, proper motions, and radial velocities to the Galactocentric reference frame. This assumed the solar velocity vector $(U_{\odot}\,,V_{\odot}\,,W_{\odot})=(11.1\,,12.24\,,7.25) \ {\rm km \ s^{-1}}$ \citep{sbd2010}, the velocity of the local standard of rest $V_{LSR}=220.0 \ {\rm km \ s^{-1}}$ \citep{majewski2008}, the Sun-GC distance $R_{\odot}=8.34 \ {\rm kpc}$ \citep{reid+2014}, and the galactocentric height of the Sun $z_{\odot}=20.8 \, \rm{ pc}$ \citep{bb2019}. We used \textsc{OrbIT} to reconstruct backwards in time the orbits of the three Cepheids for $50 \, {\rm Myr}$ with a timestep of $10^2 \, {\rm yr}$. Given the age of the Cepheids \citep[$\sim$$25$ Myr]{matsunaga+11, matsunaga+15}, the total duration of the orbital integration ensures that we have a sufficient number of complete orbits for a robust estimation of the dynamical parameters while the small timestep is needed to properly compute the orbits so deep inside the gravitational potential.

As a sanity check of our orbit integrator and of our choices for the potential model, we reproduce the results of \citet{hosek+22} by recovering the orbits for their maximum a posteriori (MAP) solutions for the Arches and Quintuplet star clusters. The results of the sanity check are in Appendix~\ref{arcqui} and show that we can effectively reproduce results already accepted in the literature.

\subsection{\textsc{OrbIT}'s outputs}\label{outputs}
\textsc{OrbIT} provides several outputs, from orbital parameters to integrals of motions (IoMs) and adiabatic invariants, the position $(X,Y,Z)$ at each timestep, and the fraction of time spent within specific radii. For the present work we focused on those that help constraining the shape and type of orbit of a tracer. The following is thus not an exhaustive list of outputs but simply the list of the variables more useful for the analysis carried out in this work.

The traditional orbital parameters, apocentre ($R_{apo}$), pericentre ($R_{peri}$), eccentricity ($ecc$), and maximum excursion from the Galactic plane ($\lvert Z_{max} \rvert$), provide important information on the region inhabited by a tracer's orbit. One of the advantages of our code is the fact that it computes a value for $R_{apo}, \, R_{peri}, \, {\rm and} \, ecc$ for each complete radial excursion of a tracer, and then provides, as a final output, the mean value of each of these parameters, together with the standard deviation of the distribution. The latter helps to quantify how well constrained each parameter is. This method allows us to account for the time-variability of the potential induced by the presence of the rotating bar of the MW. To complement the spatial information coming from the orbital parameters, we also computed the percentage of time the Cepheids spent within $300 \, {\rm pc}$ from the GC \citep[i.e. within the NSD zone of influence, ][]{sormani+22}.

In order to characterise the dynamical properties of the orbits studied, we also analysed the projected action space map, $J_{\parallel}-J_{\perp}$. These two parameters are derived from the actions, with $J_{\parallel}$ being the normalised $z$-component of the angular momentum, while $J_{\perp}$ measures the importance of radial and vertical motions \citep{vasiliev2019, naidu+2020, malhan+2022}. Following \citet{bt2008}, \textsc{OrbIT} first computes the actions  $(J_R,J_{\theta},J_{\phi})$:
\begin{equation}\label{Jphi}
    J_{\phi} = L_{z} \ ,
\end{equation}
\begin{equation}\label{Jtheta}
    J_{\theta} = L - \lvert L_z \rvert \ ,
\end{equation}
\begin{equation}\label{Jr}
    J_R = \frac{2}{\pi}\int_{R_{peri}}^{R_{apo}} \sqrt{2E-2\widetilde{\Phi}(r)-\frac{L^2}{r^2}}\rm{d}r \ ,
\end{equation}
where $E$ is the initial energy of the tracer, $L$ its total angular momentum, and $\widetilde{\Phi}(r)$ is a numerical approximation of the full potential (including all components). The approximation minimises the differences with the corresponding analytical potential when taking the mean of all axial components at distance $r$. The integral is solved numerically using a composite Simpson's 1/3 rule \citep{atkinson1991} and the results are comparable with those derived from the St\"{a}ckel-Fudge method \citep[as shown in][]{DeLeo+26_bar}.

$J_{\parallel} \, {\rm and} \, J_{\perp}$ are then derived as follows:
\begin{equation}
    J_{tot} = J_R + J_{\theta} + \lvert J_{\phi} \rvert \ ,
\end{equation}
\begin{equation}
    J_{\perp} = \frac{J_{\theta}-J_R}{J_{tot}} \ ,
\end{equation}
\begin{equation}
    J_{\parallel} = \frac{J_{\phi}}{J_{tot}} \ .
\end{equation}

Finally, the last dynamical parameter relevant to our analysis is the orbital circularity $\eta$, which is the ratio of a tracer's angular momentum to that of an hypothetical maximally rotating planar orbit with the same specific energy as the tracer \citep{abadi+2003, ruchti+2014, orkney+2023}:
\begin{equation}
    \eta = \frac{L_z}{L_z^{max}(E)} \ ,
\end{equation}
where $L_z^{max}(E)$ depends on the $(V_c, \, R_c)$ pair that minimises the equation for the energy of the tracer:
\begin{equation}
    \bigg\lvert \ \frac{V_c^2}{2} + \Phi(R_c,0) - E \ \bigg\rvert \ ,
\end{equation}
with $V_c$ the circular velocity:
\begin{equation}
    V_c = \sqrt{R_c \cdot \frac{\delta \Phi}{\delta R}\bigg\rvert_{R_c}} \ .
\end{equation}

The orientation of the Galactocentric reference frame axes adopted in this work implies that prograde tracers have negative $L_z, \, J_{\phi}, \, J_{\parallel}, \, {\rm and} \, \eta$, whereas retrograde tracers have positive values of these parameters.

\subsection{Accounting for observational errors and model dependencies}\label{obserrs}

Given the relatively small size of the NSD with respect to Galactic scales, it is important to gauge the precision of the recovered results before making any particularly strong claim. For this reason, we studied the impact of the observational errors on our results by producing, for each Cepheid, one thousand copies with phase-space information extracted from normal distributions having the observed uncertainties as $\sigma$. For this exercise, the errors on the on-sky position of the Cepheids are negligible, the errors on radial velocities are reported in \citet{matsunaga+13,matsunaga+15}, while, for the PMs, we adopted the errors from \citet{shahzamanian+22}. Finally, we allow for a $\pm 500 \, {\rm pc}$ error on the line-of-sight distances following Sec. 4 of \citet{matsunaga+15}. Throughout the paper, the sample of one thousand copies extracted from the error distribution is called the \textquote{extended sample}, while the single realisation generated from the observed parameters is called the \textquote{nominal case}.

Similarly to observational errors, there are other possible sources of systematic uncertainties in the results. The Solar phase-space values used for the coordinate transformation to the Galactocentric reference frame and the details of the adopted gravitational potential model can influence the results of our analysis. We ran a series of tests to gauge the impact of these systematic effects on the final results of our analysis for the Cepheids. We tested changing the mass of the NSD from $1.05\times10^9 \, M_{\odot}$ to $1.4\times10^9 \, M_{\odot}$ \citep{hosek+22}, changing the Sun-GC distance from $8.34 \, {\rm kpc}$ to $8.178 \, {\rm kpc}$ \citep{gravity2019}, removing the bar, or removing the spheroidal in the bulge. As a general note, the biggest impact we noted was due to changes in the mass enclosed within the region spanned by the orbits and due to the adoption of a different Sun-GC distance. Notable effects of these tests on the results for the individual Cepheids are noted in the text throughout section ~\ref{results} and the differences in the recovered orbital parameters are shown in Appendix~\ref{changes}.

\section{Results}\label{results}

As briefly anticipated in the previous section, we focused our analysis on the spatial and dynamical information that identifies the region inhabited by the orbit of each Cepheid, and the type of orbit. Generally speaking, for stars orbiting a disc-like structure (such as the NSD), we would expect prograde circular orbits. As the details of the rotation of the NSD are still debated, we do not have precise expectations regarding the defining orbital parameters of supposed NSD members, beyond expecting negative $L_z = J_{\phi}$ and $J_{\parallel} \approx \eta \leq -0.4$ (similar to the selection for Galactic disc tracers).

Given the properties of NSDs in external galaxies \citep{schultheis25}, we would also expect such orbits to be almost completely planar and to be confined within the NSD zone of influence \citep[i.e. $R_{xy} <300 \, {\rm pc}$][]{sormani+22}. These considerations translate to the expectation of low $Z_{max}$ values, and $R_{apo}$ not higher than $300 \, {\rm pc}$ at most.

For each Cepheid object of this study, we analysed the distribution of orbital parameters and the overall time spent within the NSD zone of influence. To give a clearer picture of the dynamical state of each Cepheid, we also show the orbital trajectory for $10 \, {\rm Myr}$ for both the nominal case and the full extended sample.

Moreover, dynamical studies show that stars belonging to some periodic orbital families have specific ratios between their fundamental frequencies \citep[i.e.][and references therein]{bt2008}. More specifically, the NSD has been associated with orbits belonging to the $x_2$ family (see, e.g. \citealt{Nieuwmunster+24} or \citealt{schultheis25} for a recent review) prograde orbits oscillating in the plane, perpendicularly to the major axis of the main bar and with a frequency ratio of radial and azimuthal motions of $\Omega_R:\Omega_{\phi}=2:1$ \citep{v2016}. The frequency analysis technique allows to recover the fundamental orbital frequencies from the Fourier transform of the spatial coordinate time series \citep{l1990, dl1993, bt2008, hk2009}. We used the \texttt{SuperFreq} \citep{pw2015} implementation of this technique, which employs a Fast Fourier transform, to derive the fundamental frequencies of the Cepheids' orbits in the reference frame corotating with the MW bar.

\subsection{GCC-a}\label{ceph-a}
\begin{figure}
   \centering
     \includegraphics[width=8.5cm]{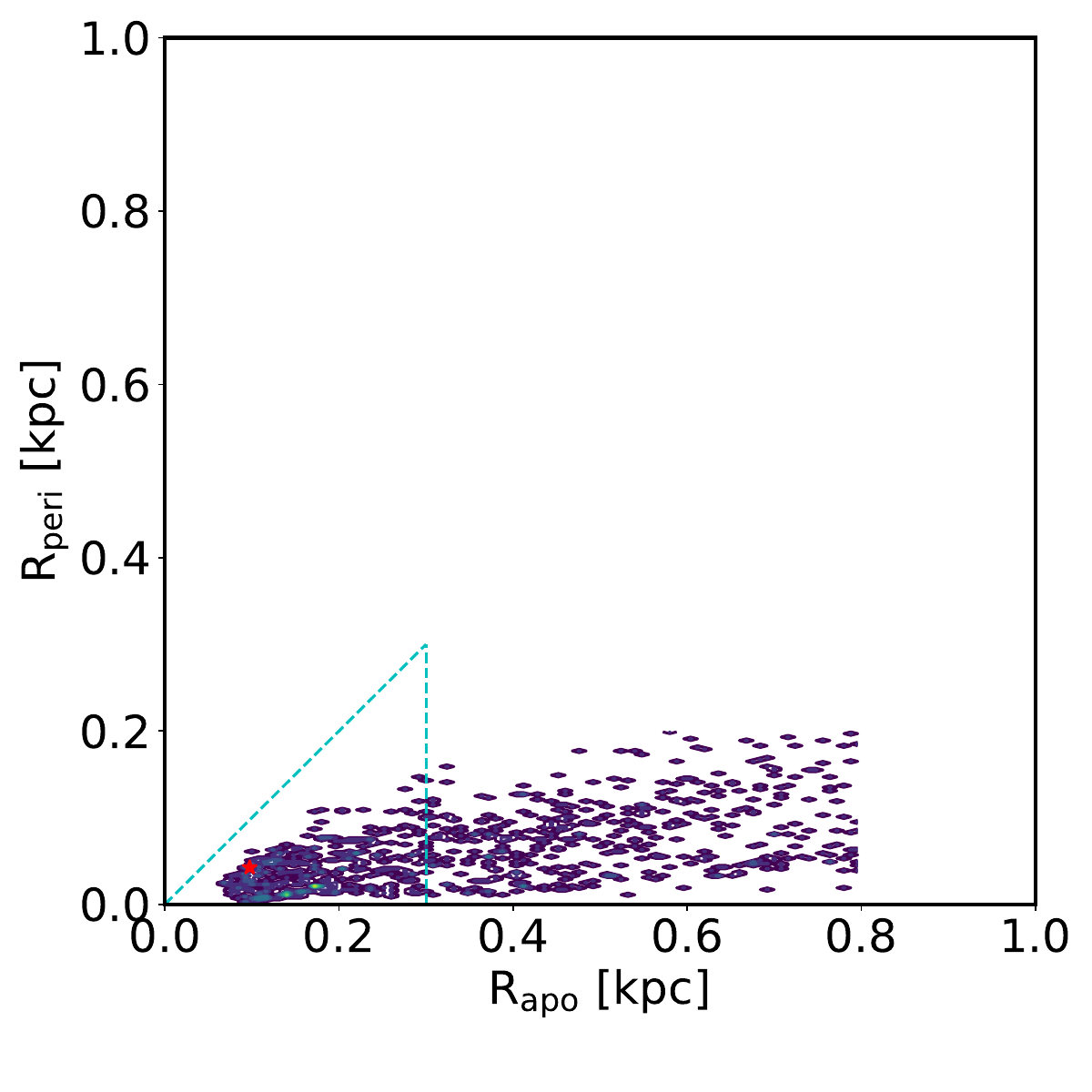}
    \caption{Distribution of the extended sample of GCC-a (density map) in the $R_{apo}-R_{peri}$ plane. The dashed cyan line identifies the NSD region (i.e. $R \leq 300 \, {\rm pc}$) while the red star is the nominal case.}\label{GCC-a_ap}
\end{figure}

The first Cepheid analysed, GCC-a \citep[number 32 in][]{matsunaga+13} is the closest to the GC. Figure ~\ref{GCC-a_ap} shows the distribution of the apocentres and pericentres of GCC-a for both the extended sample (density map) and the nominal case (red star). In the nominal case, GCC-a spends the whole time within the inner $300 \, {\rm pc}$ (the dashed cyan line), which means it stays confined within the NSD zone of influence. Nonetheless, when observational errors are taken into account, only $41.37\%$ of the extended sample spend the majority of their orbital history within the same region.

\begin{figure}
   \centering
     \includegraphics[width=8.5cm]{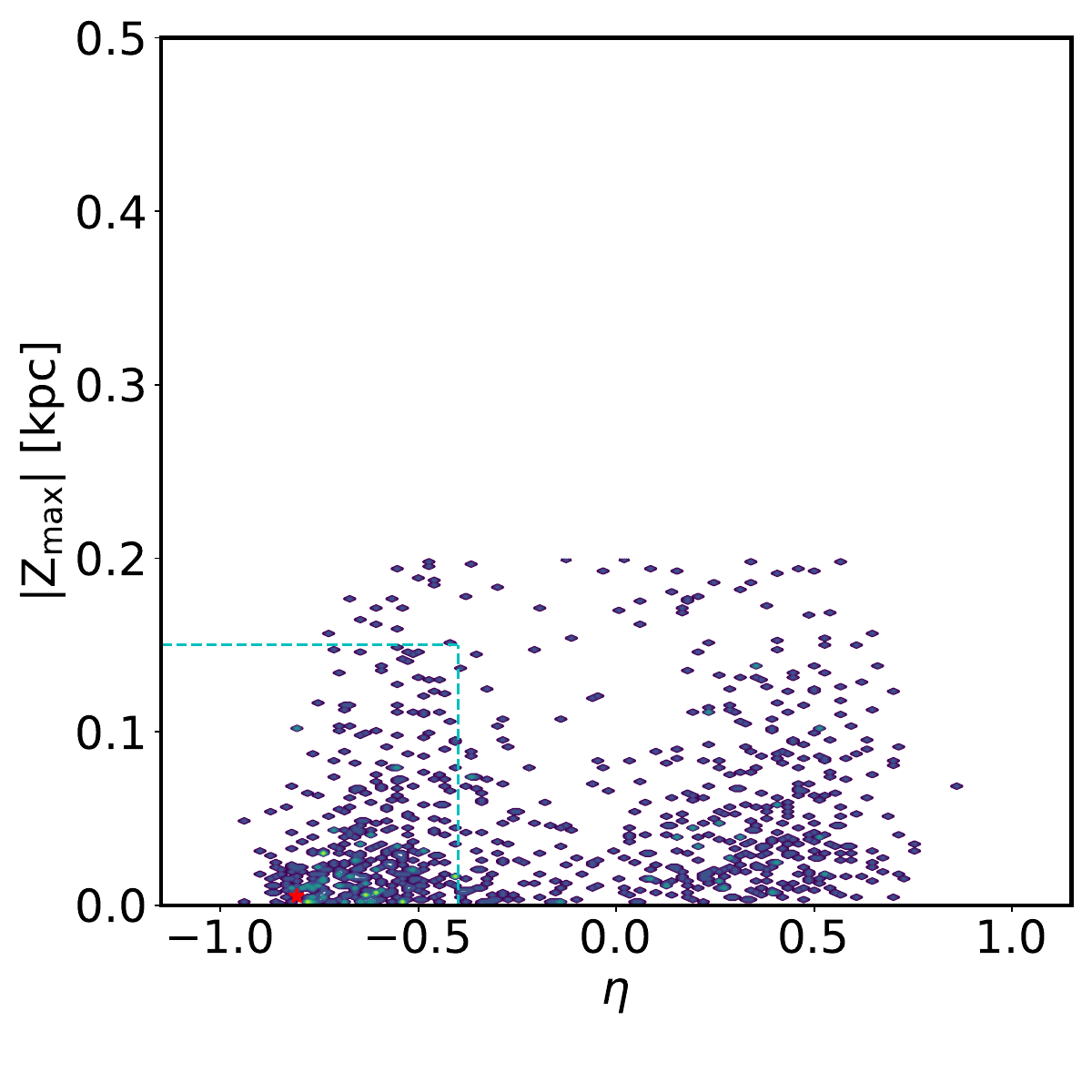}
    \caption{Distribution of the extended sample of GCC-a (density map) in the $Z_{max}-\eta$ plane. The dashed cyan line identifies the NSD region with disc-like kinematics (i.e. $Z_{max} \leq 150 \, {\rm pc}$ and $\eta \leq -0.4$) while the red star is the nominal case.}\label{GCC-a_ze}
\end{figure}

Figure ~\ref{GCC-a_ze} presents the distribution of the nominal case and extended sample in the $\lvert Z_{max} \rvert - \eta$ space. The nominal case shows a very low $\lvert Z_{max} \rvert$ and a definitely prograde orbit, while the distribution of the extended sample is more scattered. Notably, the uncertainties (chiefly the error on the distance estimation) produce a bimodal distribution in $\eta$. This is because a given value of the longitude PM results in a prograde/retrograde orbit depending on whether the Cepheid distance is smaller/larger than the distance to the GC. 

\begin{figure}
   \centering
     \includegraphics[width=8.5cm]{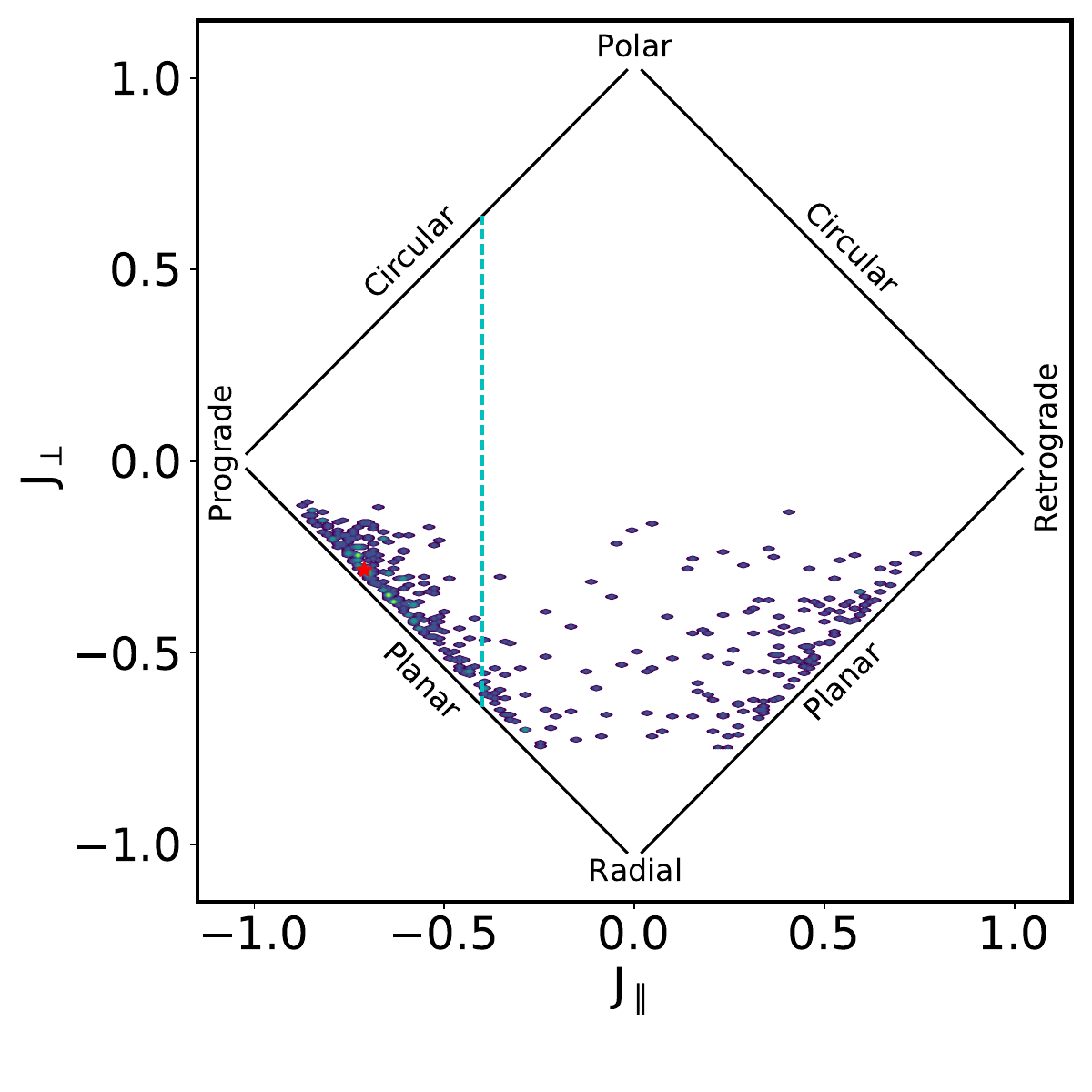}
    \caption{Distribution of a selection of the extended sample of GCC-a (density map) in the $J_{\parallel}-J_{\perp}$ plane. The selection includes only the members that spend $\geq80\%$ of their orbital history in the inner $300 \, {\rm pc}$. The dashed cyan line marks the transition to the region with disc-like kinematics (i.e. $J_{\parallel} \leq -0.4$) while the red star is the nominal case.}\label{GCC-a_jj}
\end{figure}

Figure ~\ref{GCC-a_jj} shows the nominal case and extended distribution in the projected action space, $J_{\parallel}-J_{\perp}$. Since in this dynamical space the spatial information is lost, we plot here only the orbit realisations for which the star spends most ($\geq80\%$) of its orbital history within the inner $300 \, {\rm pc}$).  In agreement with Fig.~\ref{GCC-a_ze}, GCC-a appears to be on a prograde planar orbit and it sits on the region of disc-like dynamics (left of the dashed cyan line).

For this first Cepheid, we could verify that the information yielded by Figures ~\ref{GCC-a_ap},  ~\ref{GCC-a_ze}, and ~\ref{GCC-a_jj} is fully consistent, allowing us to draw a coherent picture. Since the $Z_{max}-\eta$ plane carries inherently less information than either of the other two spaces, we will omit this plane in the remainder of the paper. We nevertheless show in Appendix~\ref{zmax} the agreement of the information gleaned from the $Z_{max}-\eta$ plane with our conclusions for the other Cepheids studied.

\begin{figure}
   \centering
     \includegraphics[width=8.5cm]{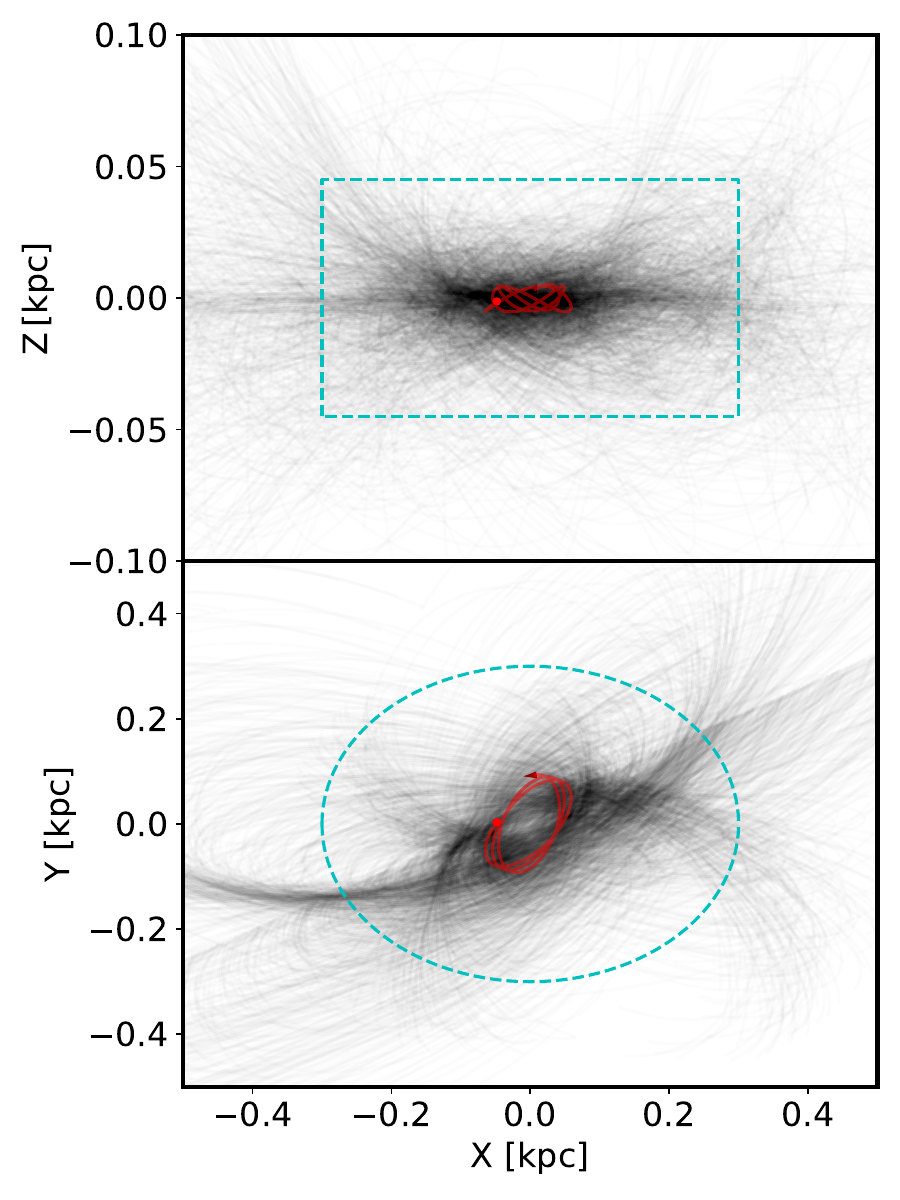}
    \caption{The orbit of Cepheid GCC-a in the XY (bottom) and XZ (top) Galactocentric reference frame corotating with the MW bar. The orbits are computed for 50 Myr but only 10 Myr are shown. The red line is the orbit of the nominal case, from the position today (red circle) to the position 10 Myr ago (arrow). The shaded black lines are the orbits of the extended sample, whereas the dashed cyan lines indicate the NSD extent.}\label{GCC-a_xyz}
\end{figure}

Figure ~\ref{GCC-a_xyz} shows the orbit of GCC-a projected on the XY (bottom panel) and XZ (top) Galactocentric planes, in the frame corotating with the bar. While accounting for the uncertainties can produce wider orbits spanning large regions of space beyond the NSD (represented by the dashed cyan lines), the nominal case (red line) stays on a planar orbit very close to the GC. Furthermore, the nominal orbit of GCC-a is almost perpendicular to the main axis of the MW bar (which is aligned to the X axis), it is a prograde orbit (it lies at negative $J_{\parallel} \, {\rm and} \, \eta$, as shown in figures ~\ref{GCC-a_ze} and ~\ref{GCC-a_jj}) and has $\Omega_R : \Omega_{\phi}=1.96$, within the customarily accepted $0.1$ tolerance interval \citep{portail2015,queiroz+21} from the ratio for $x_2$ orbits ($2:1$). Therefore, we can tentatively conclude that Cepheid GCC-a is a genuine member of the NSD, although, unfortunately, the observational errors are still too large to allow us to exclude the possibility that this is not the case, even within 1 sigma.

The tests we carried out regarding the impact of model systematics on the recovered results (see sec.~\ref{obserrs} and App.~\ref{changes}) showed that the orbit of GCC-a is the most sensitive to distance errors. This is already evident from Figures ~\ref{GCC-a_ze} and ~\ref{GCC-a_jj}, where the distribution of the extended sample shows a degree of symmetry between the prograde ($<0$) and retrograde ($>0$) regions of space. This is further reinforced by our test with a different Sun-GC distance where changing from the value provided by \citet{reid+2014} to the one from \citet{gravity2019} causes the nominal orbit of GCC-a to become retrograde (see Fig.~\ref{app-testsjj}).

\subsection{GCC-b}\label{ceph-b}

\begin{figure}
   \centering
     \includegraphics[width=8.5cm]{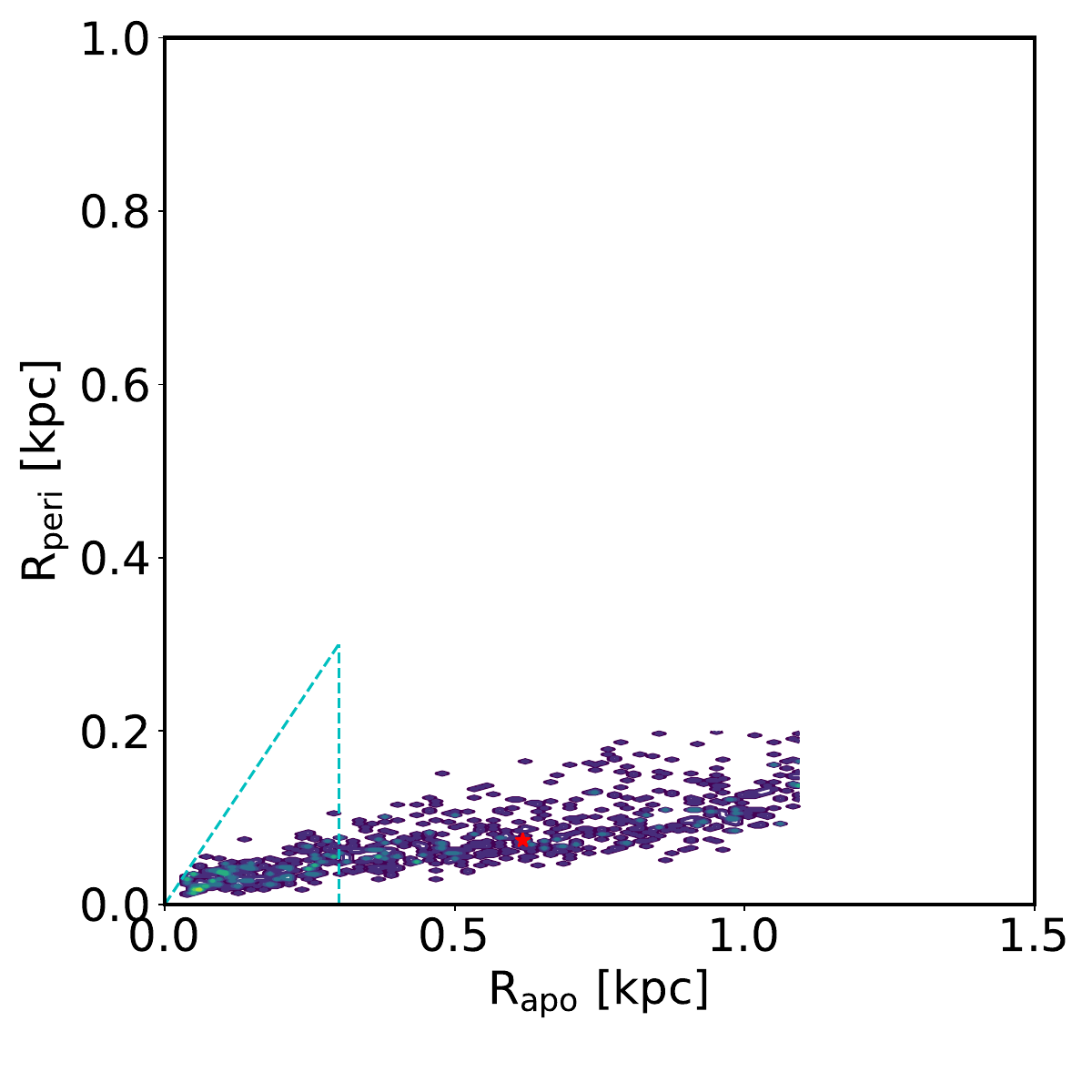}
    \caption{Same as figure ~\ref{GCC-a_ap} but for GCC-b.}\label{GCC-b_ap}
\end{figure}

\begin{figure}
   \centering
     \includegraphics[width=8.5cm]{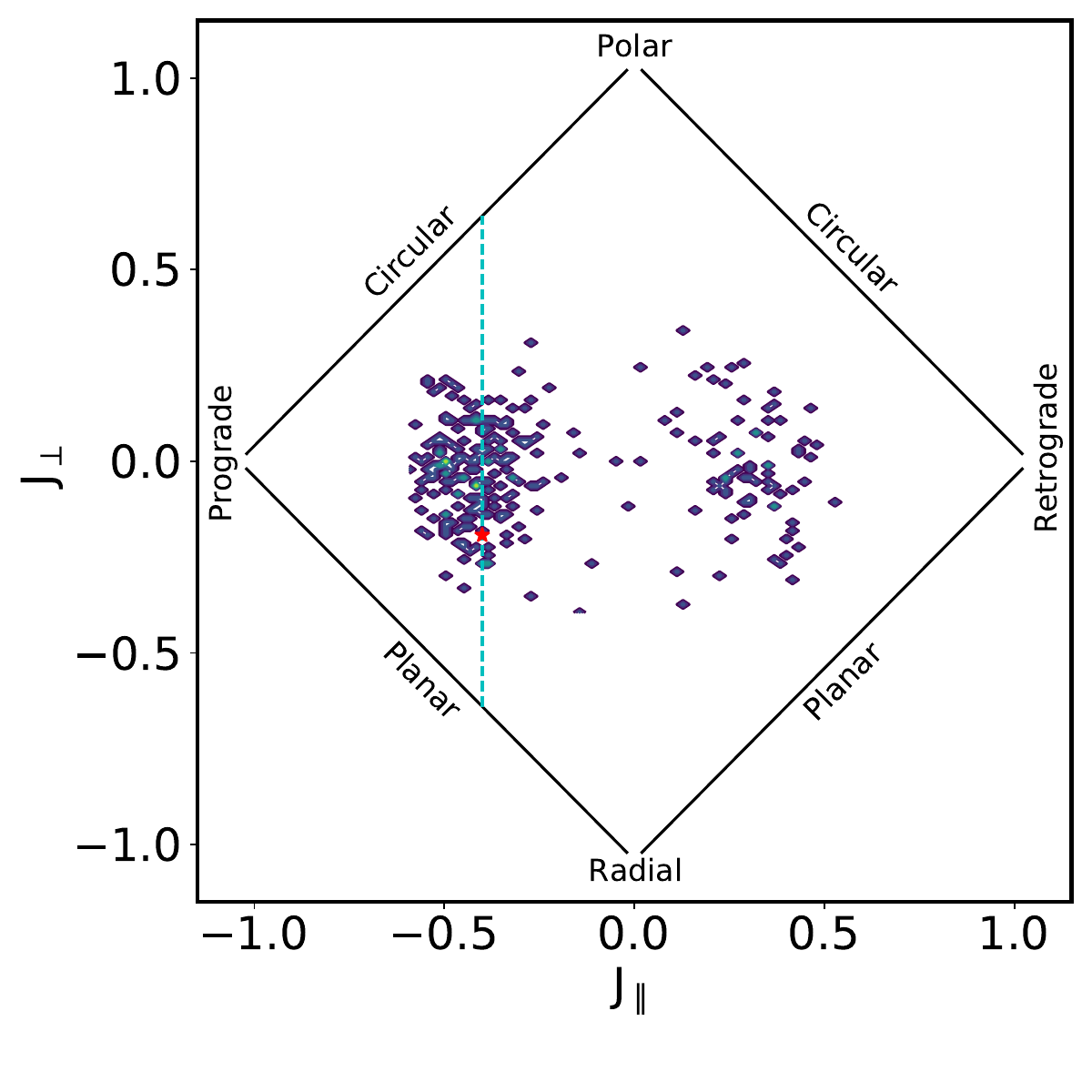}
    \caption{Same as figure ~\ref{GCC-a_jj} but for GCC-b.}\label{GCC-b_jj}
\end{figure}

\begin{figure}
   \centering
     \includegraphics[width=8.5cm]{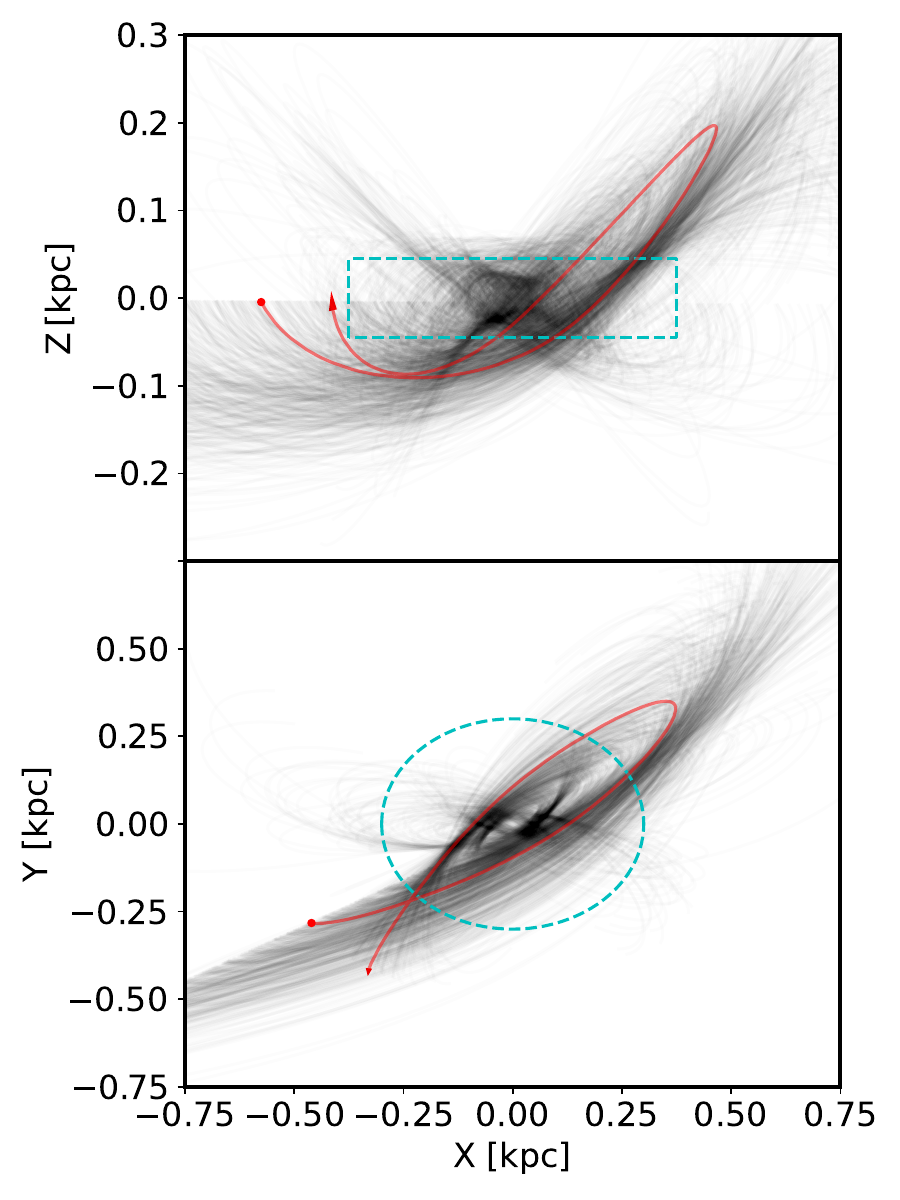}
    \caption{Same as figure ~\ref{GCC-a_xyz} but for GCC-b.}\label{GCC-b_xyz}
\end{figure}

Following the same analysis done for GCC-a, we present the nominal case (red star) and extended sample (density map) for GCC-b in the spaces of $R_{apo}-R_{peri}$ (Fig.~\ref{GCC-b_ap}), $J_{\parallel}-J_{\perp}$ (Fig.~\ref{GCC-b_jj}), and the orbits in the XY and XZ planes (bottom and top panels of Fig.~\ref{GCC-b_xyz}, respectively). Fig.~\ref{GCC-b_ap} shows that GCC-b preferentially inhabits regions of space outside the putative size of the NSD (dashed cyan lines), spending only $28.27\%$ of its orbital time within the inner $300 \, {\rm pc}$ in the nominal case, and with only $28.73\%$ of the extended sample spending the majority (80$\%$) of their time within the same region.

Fig. ~\ref{GCC-b_jj} shows unconstraining dynamical information as the nominal case sits squarely on the dashed cyan line that delimits the zone of disc-like kinematics ($J_{\parallel} \leq -0.4$) and the majority of the extended sample also lies in this transitional zone.

The orbits presented in Fig.~\ref{GCC-b_xyz} corroborate the results from Fig.~\ref{GCC-b_ap}. GCC-b has an orbit that crosses quickly the NSD region (dashed cyan line) with a maximum Z excursion about four times larger that the NSD height ($200 \, {\rm pc}$ versus $\sim 50 \, {\rm pc}$). Even taking into account the observational errors does not help in confining the orbits within the NSD, with a majority of realisations of the extended sample (shaded black lines) populating wide orbits. The nominal orbit of GCC-b is prograde but it is too wide, not perpendicular to the bar main axis and it has $\Omega_R : \Omega_{\phi}=1.54$. For these reasons the orbit of GCC-b does not belong to the $x_2$ family.

Our tests showed that GCC-b is the most robust against systematic effects. This Cepheid keeps a prograde orbit that lies mostly outside the NSD region in all cases, becoming slightly more circular and prograde in the test without the spheroidal component of the bulge (i.e. with much less mass enclosed by the orbit).

\subsection{GCC-c}\label{ceph-c}

\begin{figure}
   \centering
     \includegraphics[width=8.5cm]{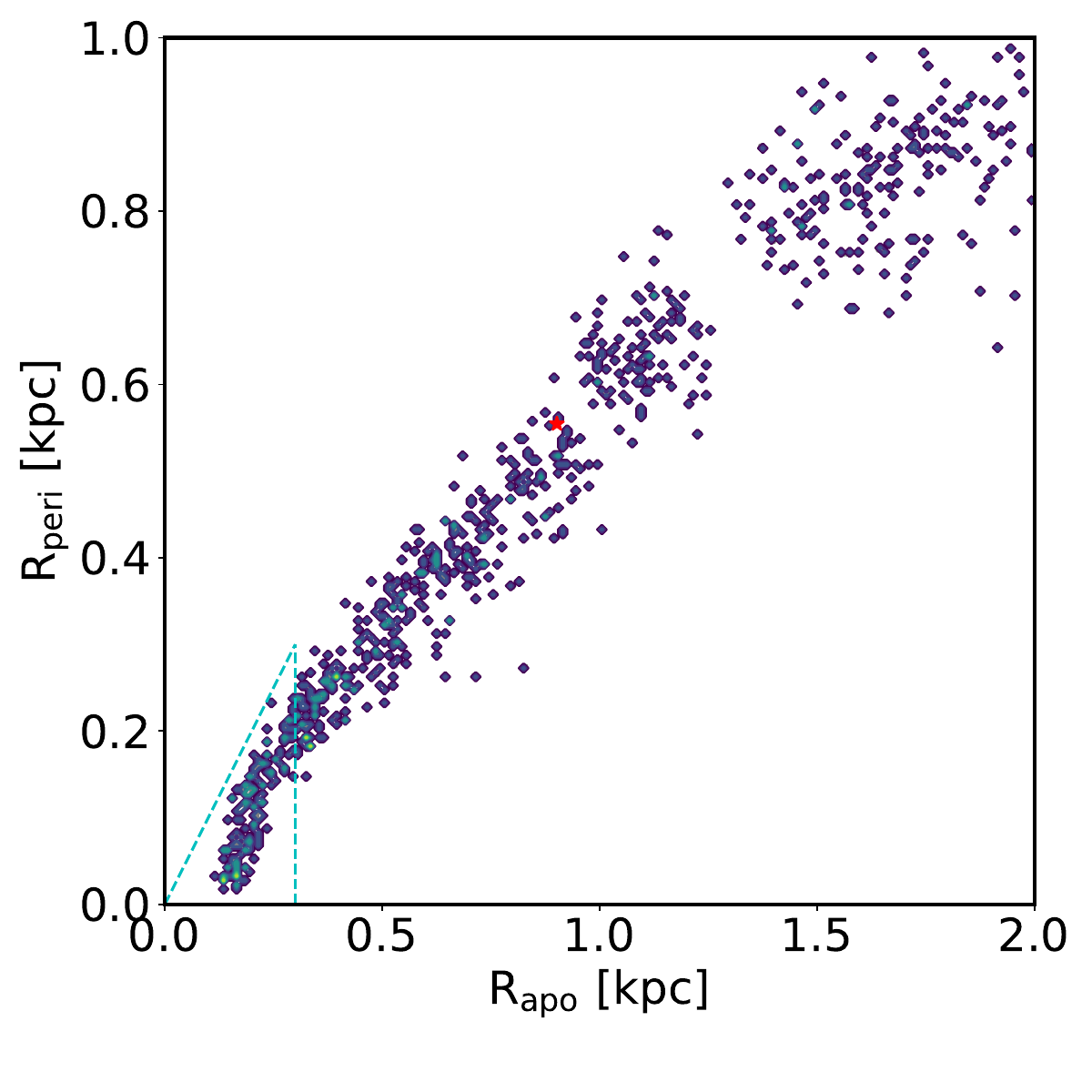}
    \caption{Same as figure ~\ref{GCC-a_ap} but for GCC-c.}\label{GCC-c_ap}
\end{figure}

GCC-c \citep[number 18 in][]{matsunaga+13} is the Cepheid more distant from the GC. From Fig.~\ref{GCC-c_ap} it appears immediately clear that it has an orbit much wider than the NSD size (dashed cyan lines), with a nominal $R_{peri} \, {\rm of} \, 555 \, {\rm pc}$ (red star) and the distribution of the extended sample (density map) spanning a wide range of values.

\begin{figure}
   \centering
     \includegraphics[width=8.5cm]{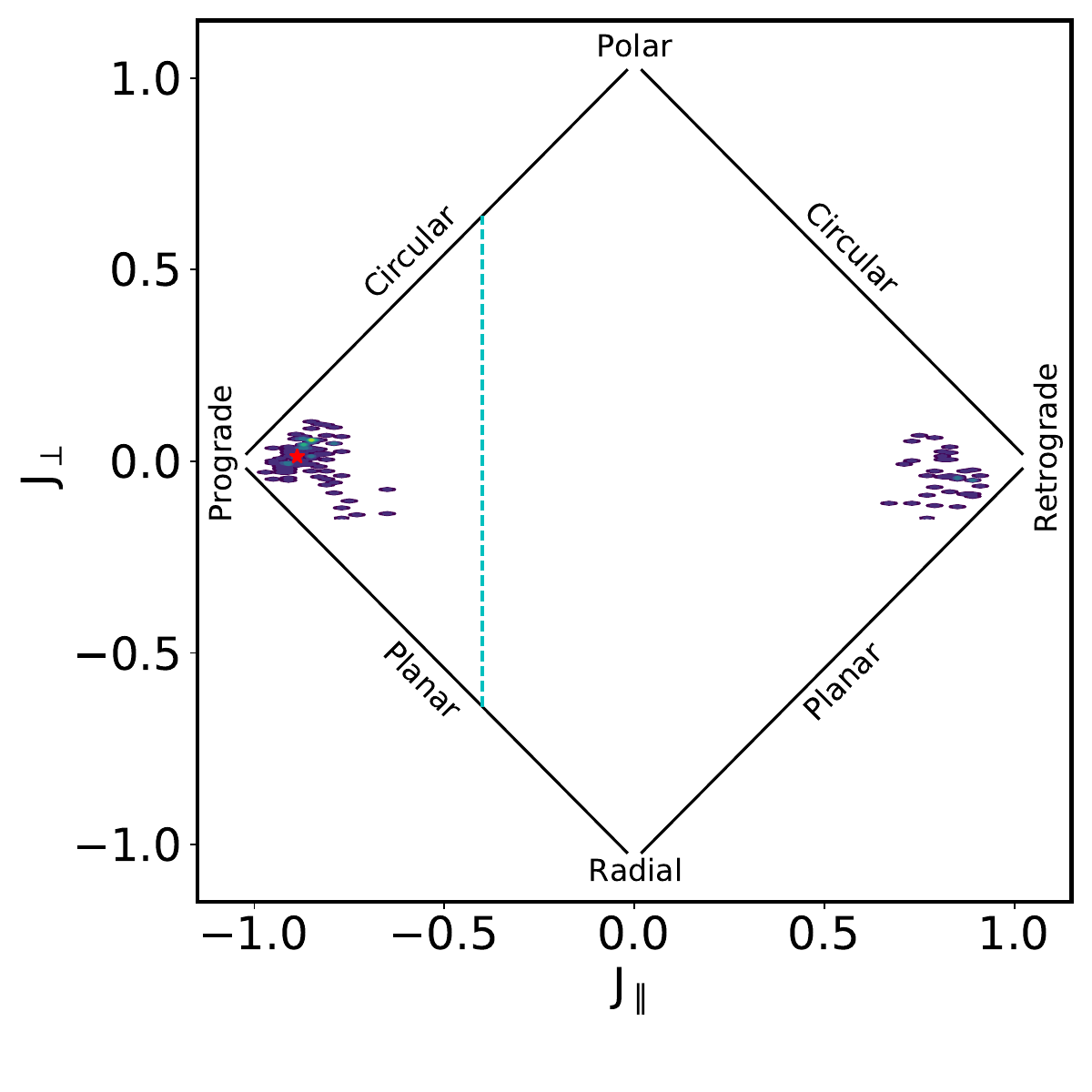}
    \caption{Same as figure ~\ref{GCC-a_jj} but for GCC-c.}\label{GCC-c_jj}
\end{figure}

In Fig.~\ref{GCC-c_jj} we show the distribution of the nominal case and extended sample for GCC-c in $J_{\parallel}-J_{\perp}$ space. This Cepheid has the most prograde and rotational supported orbit, with both the nominal case (red star) and extended sample (density map) clustered around $J_{\parallel} \sim -0.88$.

\begin{figure}
   \centering
     \includegraphics[width=8.5cm]{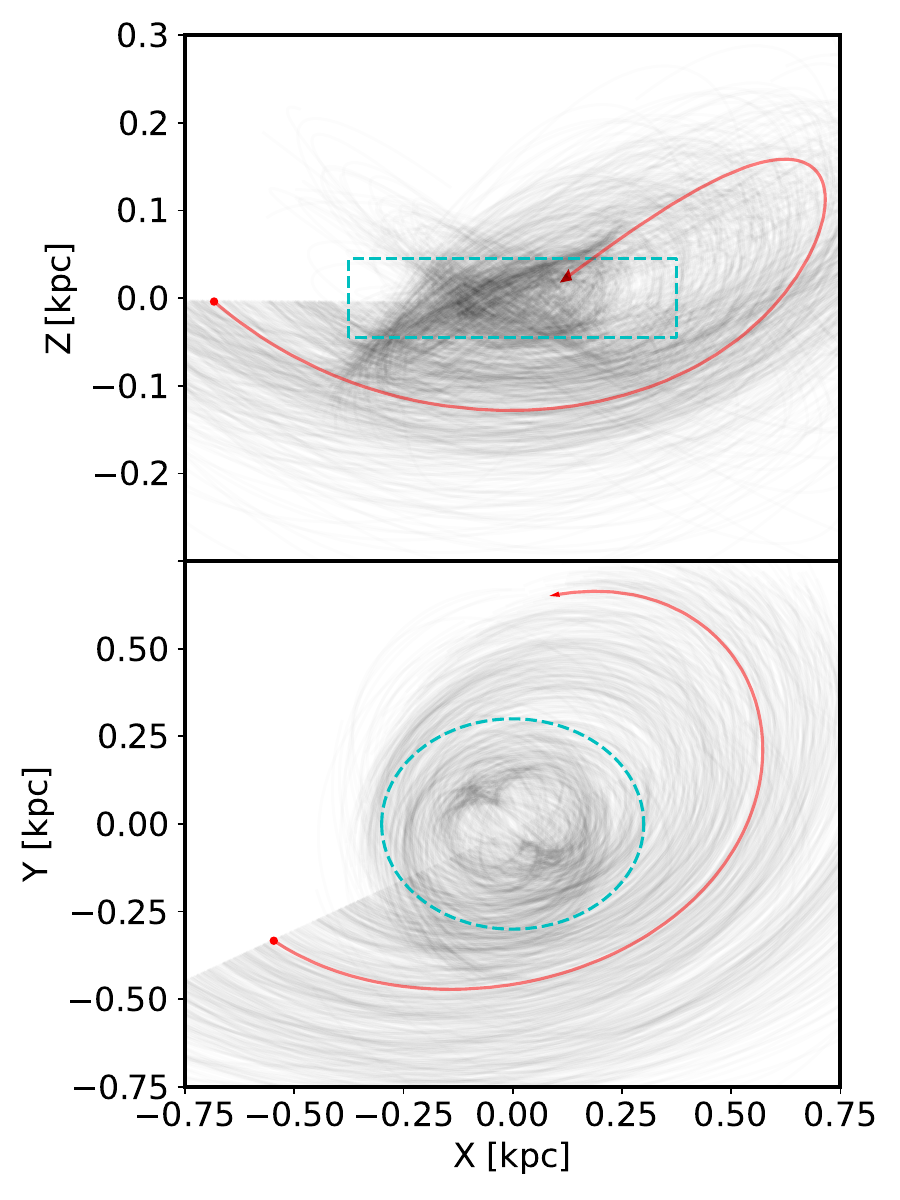}
    \caption{Same as figure ~\ref{GCC-a_xyz} but for GCC-c.}\label{GCC-c_xyz}
\end{figure}

The wide orbits of GCC-c (Fig.~\ref{GCC-c_xyz}) lie well outside the NSD. The nominal case (red line) never enters the inner $300 \, {\rm pc}$ and only $23.45\%$ of the members of the extended sample (shaded black lines) spend the majority of their orbital history in the NSD region. The nominal orbit of GCC-c is extremely prograde, wide and disc-like, and it has $\Omega_R : \Omega_{\phi}=1.37$, thus it is not an $x_2$ orbit.

Tests with different model assumptions have shown that GCC-c is quite resilient to changes due to most systematic effects. While it is not impacted by the removal of the bar or by changes in the Sun-GC distance, the removal of the spheroidal component of the bulge has a dramatic effect on the orbit of this Cepheid. With $10^{10} \, M_{\odot}$ less mass in the inner region of the MW, GCC-c is not confined to the bulge region anymore and becomes a bona fide disc star (see Fig.~\ref{app-testsap}).

\section{Discussion and conclusions}\label{concl}

The nuclear bulge of the MW is a poorly studied region, due to high crowding and extreme interstellar extinction. Yet, it is an extremely rich region, present in every other spiral galaxy, that only in the MW can be resolved down to individual stars. It is, arguably, {\em the} most interesting square degree of the whole Galaxy. As only very recently the available instrumentation allowed us to probe into it, several open questions remains \citep[see, e.g.][for a recent review]{schultheis25}. For instance, one of the key aspects to try and clarify is the Star Formation History of this region, providing crucial information on its formation mechanism, believed to be intimately related to the formation of the main bar \citep[e.g.,][]{fragkoudi16, sormani20b, tress20, schultheis+21, safreitas25}. The derivation of the age distribution of stars in the standard way, from the CMD, is very difficult due to the superposition of different stellar components along the line of sight, strongly contaminating the main sequence turnoff region, and due to the low sensitivity of brighter features such as the RC to stellar age \citep{nogueras-lara+20}. For this reason, the independent constraint on the ages provided by the presence of pulsating stars is especially important. Classical Cepheids, in particular, are unequivocally young stars (with ages around 20-200 Myr), highlighting relatively recent star formation, and thus also informing about the close past of the gas in the CMZ. For this reason, establishing whether these 4 variables, the only Cepheids discovered in the MW nuclear region, genuinely belong to the NSD or they are just passing by, has important implications for the formation of the inner Galaxy.

It is important to note here that previously available information was insufficient to establish NSD membership. Distance and position measurements cannot rule out if a tracer is an interloper simply passing by the studied region and aligning by chance with the GC. The velocity distributions of the structures in the region have substantial overlap \citep[][and see Appendix~\ref{PMs_RV}]{shahzamanian+22}, not allowing to decisively assign individual tracers to any specific population based solely on its velocity vector. This means that for the three Cepheids object of our study, their NSD membership is an open question and is not informed, biased or constrained by any strong prior. Even considering the biases introduced by imposing a gravitational potential model (biases for which we fully account and that we discuss in Sect~\ref{obserrs} and Appendix~\ref{changes}) a complete characterisation through orbital integration is needed to get the full dynamical picture. The uncertainties involved in the observations and introduced by the adoption of a specific potential model for the analysis also do not allow to completely rule out any scenario but our study gives the first complete dynamical outlook on the issue of NSD membership for the three Cepheids studied and points to the most probable answer for each of them.

In the present paper, based on an updated MW potential and a novel method to take into account the temporal change in the orbital parameters due to the rotation of the bar, as well as the impact of the observational errors, we can investigate, for the first time in a quantitative manner, whether these Cepheids truly belong to the NSD. We conclude that, based on the nominal results, GCC-a should be a member of the NSD, GCC-b is probably not a NSD member, while GCC-c is almost certainly not a NSD member. GCC-d, on the other hand, has an uncertain classification as a classical Cepheid. It  does not have PMs measured homogeneously to the others and therefore we didn't compute its orbit and excluded it from our analysis.
Taking into account the observational errors, the situation for GCC-a appears less clear, with wider and progressively less circular orbits. GCC-b would become more prograde, but its orbit remains too wide, reaching far above the Galactic plane, with respect to the NSD size. The orbits of GCC-c are very circular but too wide in the large majority of cases, suggesting that, rather than a member of the NSD, it might be a thin disc star on an orbit very close to the GC.

In summary, the nominal solutions favour a single genuine NSD member out of three studied while the realisations sourced from the error distributions allow for a residual association for GCC-b and GCC-c but put into question the case of GCC-a. Current data, especially on the distance determinations, do not allow for a conclusive answer to the question of NSD association for these three Cepheids. Even so, our analysis shows that only one Cepheid can be robustly associated with the NSD and this would weaken rather than confirm the recent star-formation burst previously inferred for the NSD.

\begin{acknowledgements}
      MDL acknowledges financial support from the project \textquote{LEGO – Reconstructing the building blocks of the Galaxy by chemical tagging} (PI: Mucciarelli) granted by the Italian MUR through contract PRIN2022LLP8TK\_001. MDL further acknowledges support from the project GENESIS - Searching for the primordial structures of the Universe in the heart of the Galaxy (Advanced Grant FIS-2024-02056, PI:Ferraro), funded by the Italian MUR through the Fondo Italiano per la Scienza call.
      R.A. acknowledges support from the Lise Meitner grant 932 from the Max Planck Society (grant PI: M. Bergemann) and from the European 933 Research Council (ERC) under the European Union’s Horizon 2020 research 934 and innovation programme (Grant agreement No. 949173, PI: M. Bergemann).
      This work was partially funded by ANID BASAL Center for Astrophysics and Associated Technologies (CATA) FB210003, and by FONDECYT Regular grant No. 1230731.
      BAT acknowledges support from the National Agency for Research and Development (ANID) Doctorado Nacional grant 21231305/2023.
      This research made use of the Astropy \citep{astropy2013, astropy2018}, Matplotlib \citep{matplotlib} and Numpy \citep{numpy} packages.
 \end{acknowledgements}

\bibliographystyle{aa}
\bibliography{jwst}

\newpage

\begin{appendix}

\section{PMs and RVs distributions}\label{PMs_RV}

\begin{figure}[!h]
    \includegraphics[width=0.48\textwidth]{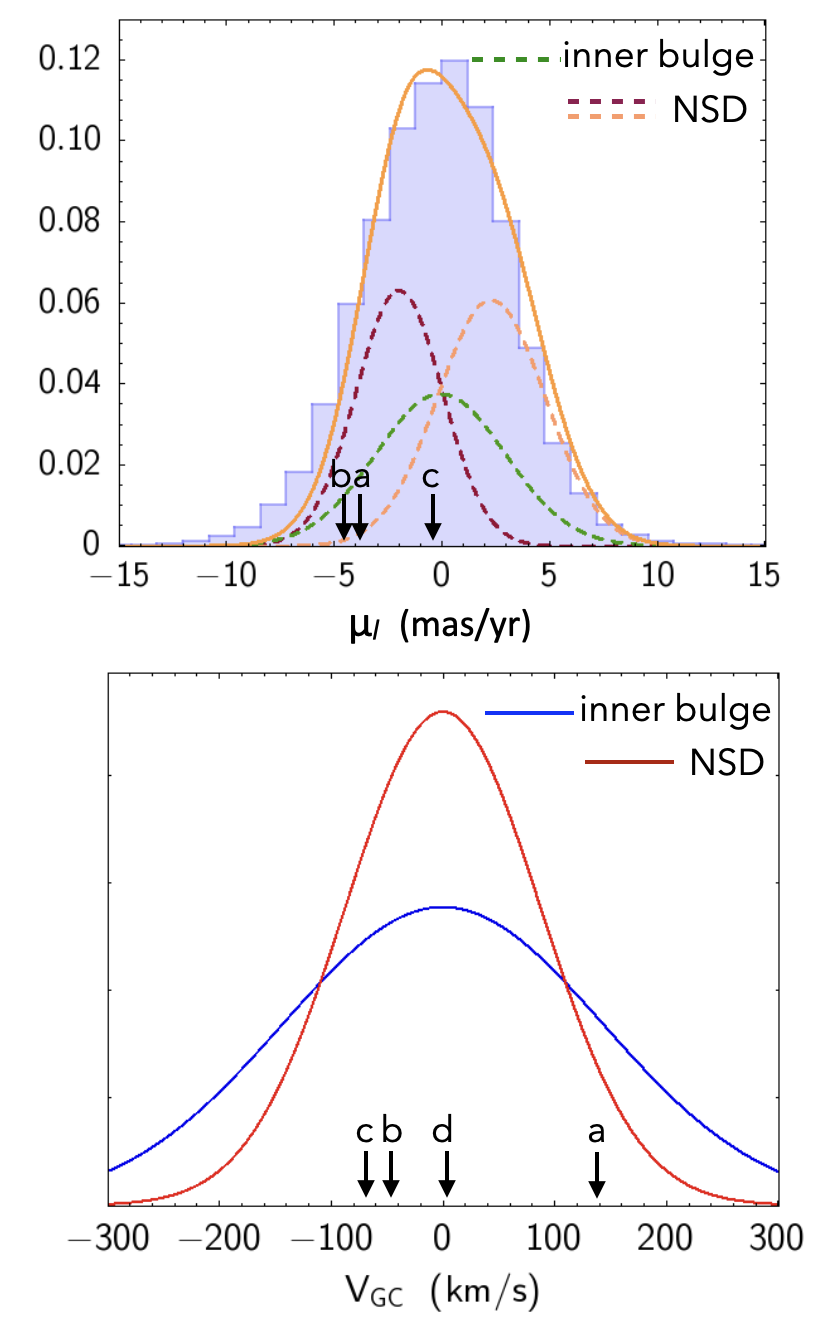}
    \caption{Cepheid kinematics parameters. Top: The longitude PM for GCC-a, GCC-b and GCC-c are shown against the longitude PM distribution of NSD stars, as in Fig.9 of \citet{shahzamanian+22}. The authors fit the observed distribution as the sum of three Gaussians: one for inner bulge stars (green) and two for NSD stars in front (red, moving eastward) or behind (orange, moving westward) the GC. Bottom: RVs for the four Cepheids are shown together with the RV distribution of NSD stars, having $\sigma_{\rm RV}$=87 km/s \citep{schultheis+21} and of inner bulge stars, with $\sigma_{\rm RV}$=144 km/s \citep{quezada25}.
}
    \label{fig:pml_RV}
\end{figure}

Figure~\ref{fig:pml_RV} shows the Cepheids (black arrows) PM in longitude (top) and RV (bottom) against the velocity distribution for inner bulge and NSD stars, as reported in the literature \citep{shahzamanian+22, schultheis+21, quezada25}. Both panels highlight the fact that the kinematic parameters alone do not allow us to properly constrain the Cepheid membership, as the distribution of different Galactic components towards the GC largely overlap. The orbit integration presented here, although limited by the current uncertainties in the measured distance, provides, for the first time, a more physically supported analysis of the Cepheid membership.

\FloatBarrier

\section{Sanity check for \textsc{OrbIT} with the NSD}\label{arcqui}

\begin{figure}[b!]
   \centering
     \includegraphics[width=8.5cm]{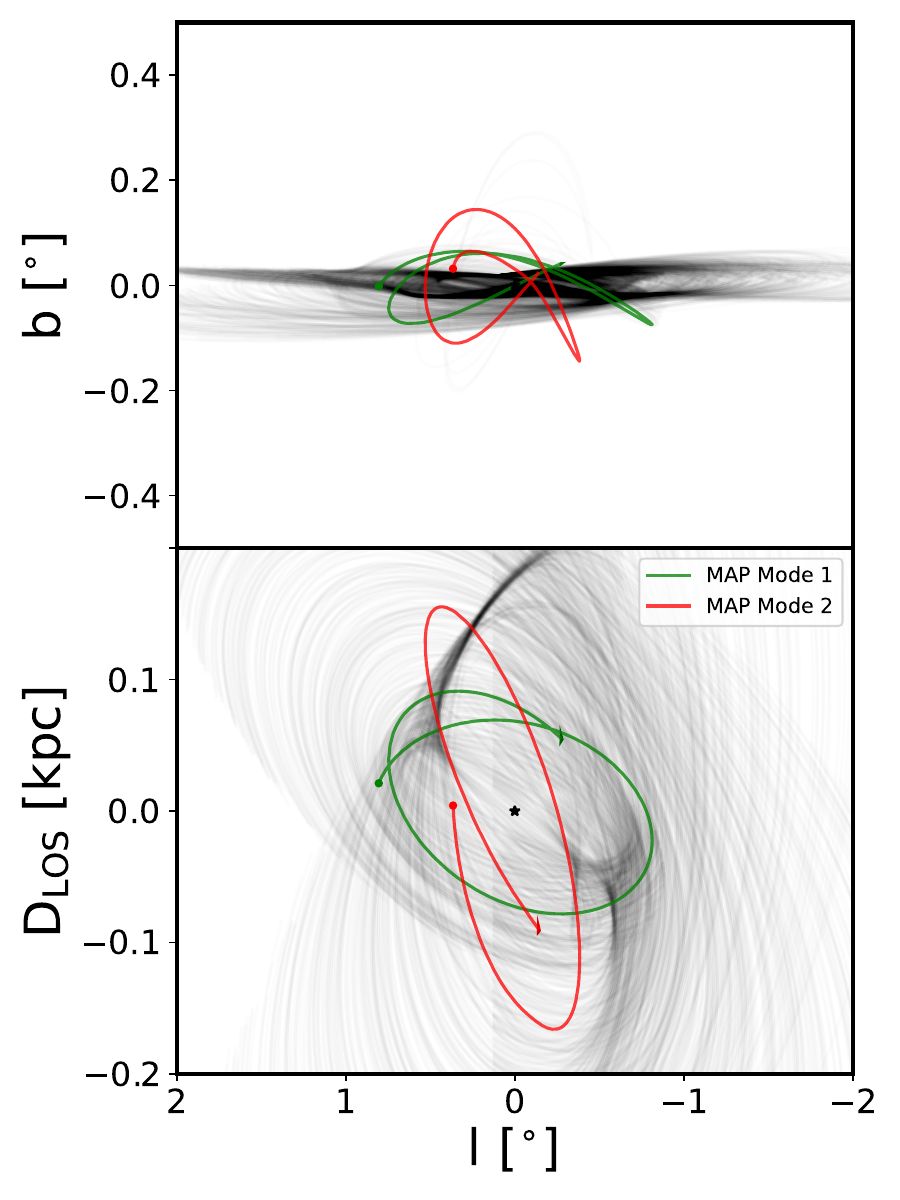}
    \caption{The orbit of the Arches cluster in the Galactic coordinates (top) and Galactic longitude vs $D_{LOS}$ (bottom). The orbits of the first 5 Myr of a 50 Myr long integration are shown, in shaded black for the 1000 realisations generated from the observational constraints, in green our forward integration for \citet{hosek+22} MAP Mode 1 and in red for MAP Mode 2. The dots mark the start of the MAP orbits while the arrows mark their direction, the black star represents the GC.}\label{app-arc}
\end{figure}

\begin{figure}
   \centering
     \includegraphics[width=8.5cm]{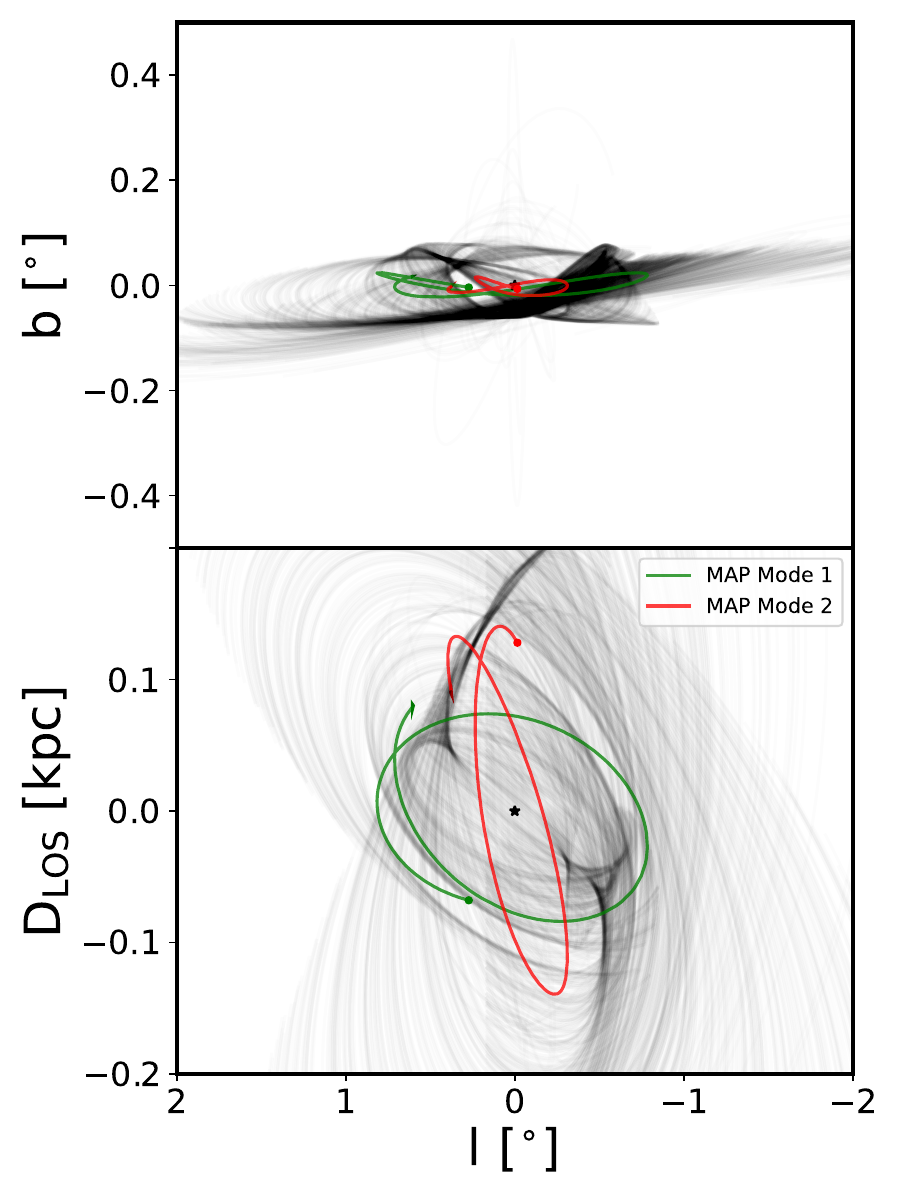}
    \caption{Same as figure ~\ref{app-arc} but for the Quintuplet cluster.}\label{app-qui}
\end{figure}

\citet{hosek+22} provides present-day observational constraints on the position and velocities in the Galactic reference frame of the Arches and Quintuplet stellar clusters (their Table 7) and also the MAPs of their birth phase-space parameters in the Galactocentric reference frame (their table 8). We used the observational constraints to generate 1000 copies of the two clusters with different initial conditions and integrate their orbits backwards for 50 Myr. We also used the MAPs solutions to integrate the clusters' orbits forward for 50 Myr. Fig.~\ref{app-arc} shows the results of our orbital integrations for the Arches cluster: the shaded black lines are the backwards integrated orbits of the 1000 realisations sourced from the observational errors, the green line is the forward integrated orbit of MAP Mode 1, while the red line is for MAP Mode 2. We managed to recover with our integrator and choice of potential model the orbits shown in Fig. 5 of \citet{hosek+22}. Fig.~\ref{app-qui} is the same for the Quintuplet cluster and the comparison with Fig.6 of \citet{hosek+22} is also successful.

\section{Impact of changes to the underlying potential model}\label{changes}

\begin{figure}
   \centering
     \includegraphics[width=8.5cm]{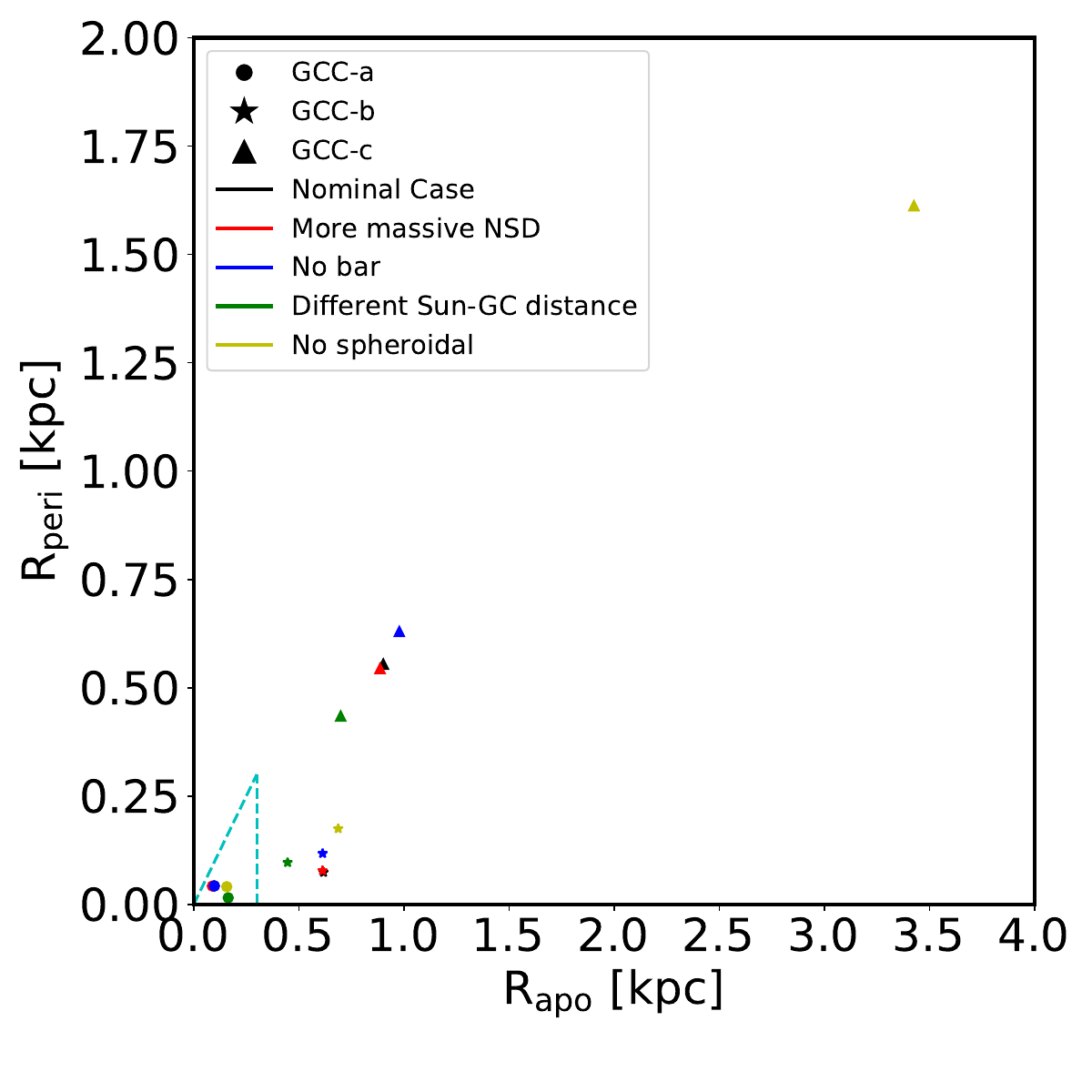}
    \caption{Same as figure ~\ref{GCC-a_ap} but for the different tests with changes on the underlying model assumptions. Different symbols mark the GCCs while the different colors are used to distinguish the various tests, as specified in the legend.}\label{app-testsap}
\end{figure}

\begin{figure}
   \centering
     \includegraphics[width=8.5cm]{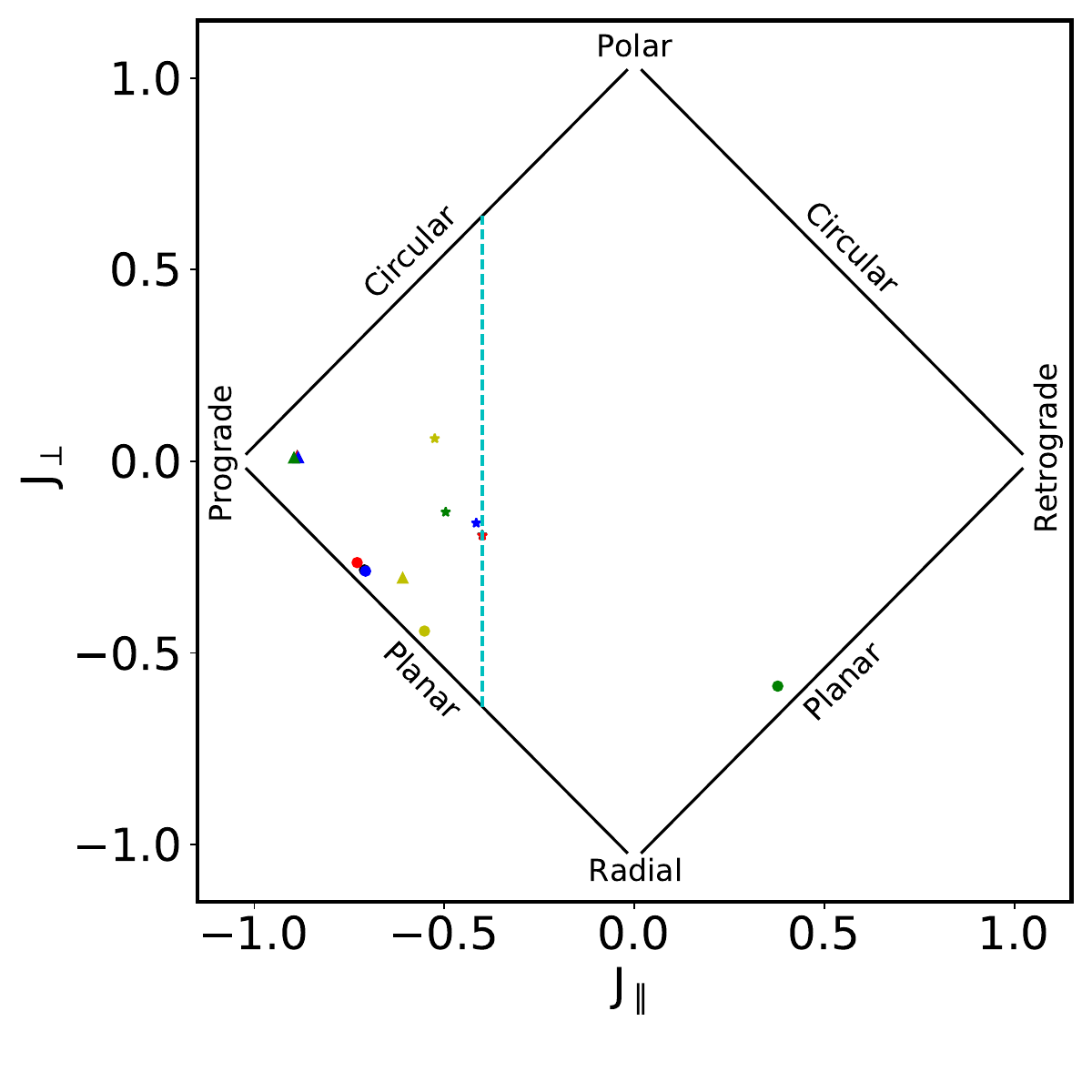}
    \caption{Same as figure ~\ref{app-testsap} but for the $J_{\parallel}-J_{\perp}$ dynamical parameter space.}\label{app-testsjj}
\end{figure}

We tested our results against different model choices, where for \textquote{model} we use here the widest definition of it being comprised of all the theoretical assumptions that are made before computing the orbits. As detailed in sec.~\ref{obserrs}, we tested the impact of a different mass distribution within the inner regions of the MW by removing the bar, the spheroidal component of the bulge and changing the mass of the NSD (each a separate test). We also changed the assumed Sun-GC distance, as distance uncertainties might result in different relative positions of the Cepheids and the GC (i.e. the stars might switch between being \textquote{foreground} and \textquote{background} with respect to the GC). Figures ~\ref{app-testsap} and ~\ref{app-testsjj} reproduce the two dynamical parameter spaces we have used throughout the paper to assess the dynamical qualities of the Cepheids, respectively $R_{apo}-R_{peri}$ and $J_{\parallel}-J_{\perp}$. In both figures, we used a symbol to denote each Cepheid and different colors for the different tests (as per the legend of fig.~\ref{app-testsap}), which means that, for example, the results for GCC-b in the test with the more massive NSD are indicated by red stars.

Fig.~\ref{app-testsap} shows the possible changes to our results in terms of the region of space occupied by the orbits of the Cepheids. GCC-a (the filled circles) is the least affected as its $R_{apo} \, {\rm and} \, R_{peri}$ change very little. The other two Cepheids show that, with less mass within the system (blue symbols for the removal of bar or yellow symbols for the removal of the spheroidal), their orbits are wider (with GCC-c the most affected, as clear from the position of the yellow triangle). Changing the Sun-GC distances (green symbols) seemed to produce the opposite effect of binding the Cepheids more. Finally, from all three stars, it seems that even a $\sim33\%$ increase of the mass of the NSD does not significantly change the results with respect to the nominal case (the red and black symbols are always close together and overlapping).

Fig.~\ref{app-testsjj} shows the dynamical content of the orbits and the changes due to our tests. Generally speaking, the removal of the bar (blue symbols) and the increase of the NSD mass (red symbols) do not seem to change the results much with respect to the nominal cases (black symbols). For GCC-a, removal of the spheroidal causes the orbit to be less prograde (yellow circle) while a change in the Sun-GC distance induces a switch to retrograde space (green circle). This is most probably due to the star now being seen as background with respect to the GC but moving in the same direction as before, and so rotating in the opposite sense than the nominal case. GCC-b becomes slightly more prograde with the different Sun-GC distance (green star) or with less mass in the inner region (yellow star), which also makes it more circular. The only notable change for the dynamics of GCC-c is the removal of the spheroidal (yellow triangle), which makes the Cepheid much less prograde and more planar.

\FloatBarrier

\section{$\lvert Z_{max} \rvert - \eta$ plane}\label{zmax}

\begin{figure}
   \centering
     \includegraphics[width=8.5cm]{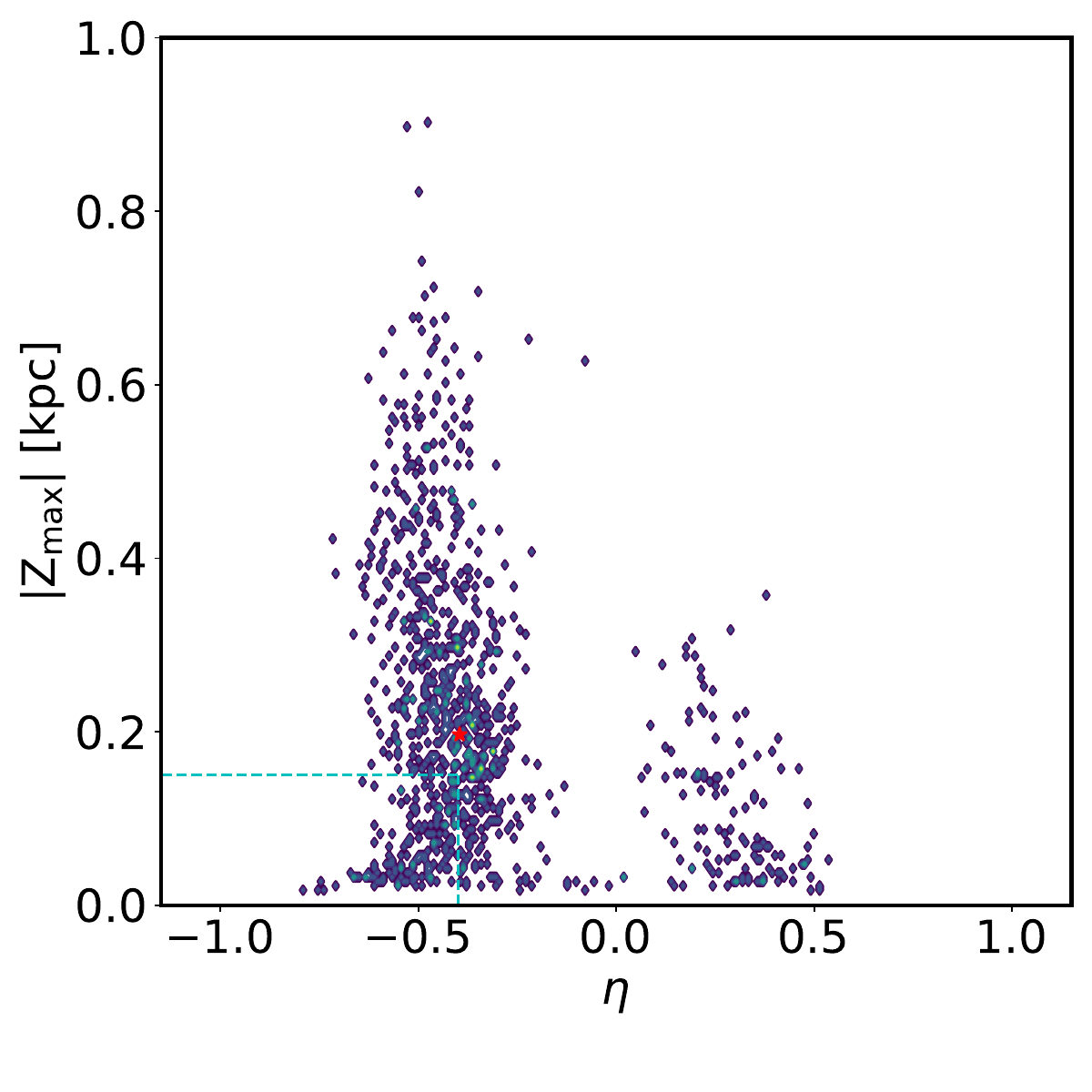}
    \caption{Same as figure ~\ref{GCC-a_ze} but for GCC-b.}\label{app-b_ze}
\end{figure}

As mentioned in Sect~\ref{ceph-a}, the $\lvert Z_{max} \rvert - \eta$ plane contains partial hybrid information on both the spatial distribution and dynamical content of an orbit. This is redundant when the same information can be taken in a more complete form from the other dynamical parameter spaces we analysed in this work. We show here that the agreement of the spatial and dynamical information that we found for GCC-a is confirmed also for GCC-b and -c. Fig.~\ref{app-b_ze} shows GCC-b in $\lvert Z_{max} \rvert - \eta$ space, the Cepheid appears to be on threshold for disc-like dynamics and inhabit a region of space outside the NSD, as we concluded in the main text (Sect.~\ref{ceph-b}).

\begin{figure}
   \centering
     \includegraphics[width=8.5cm]{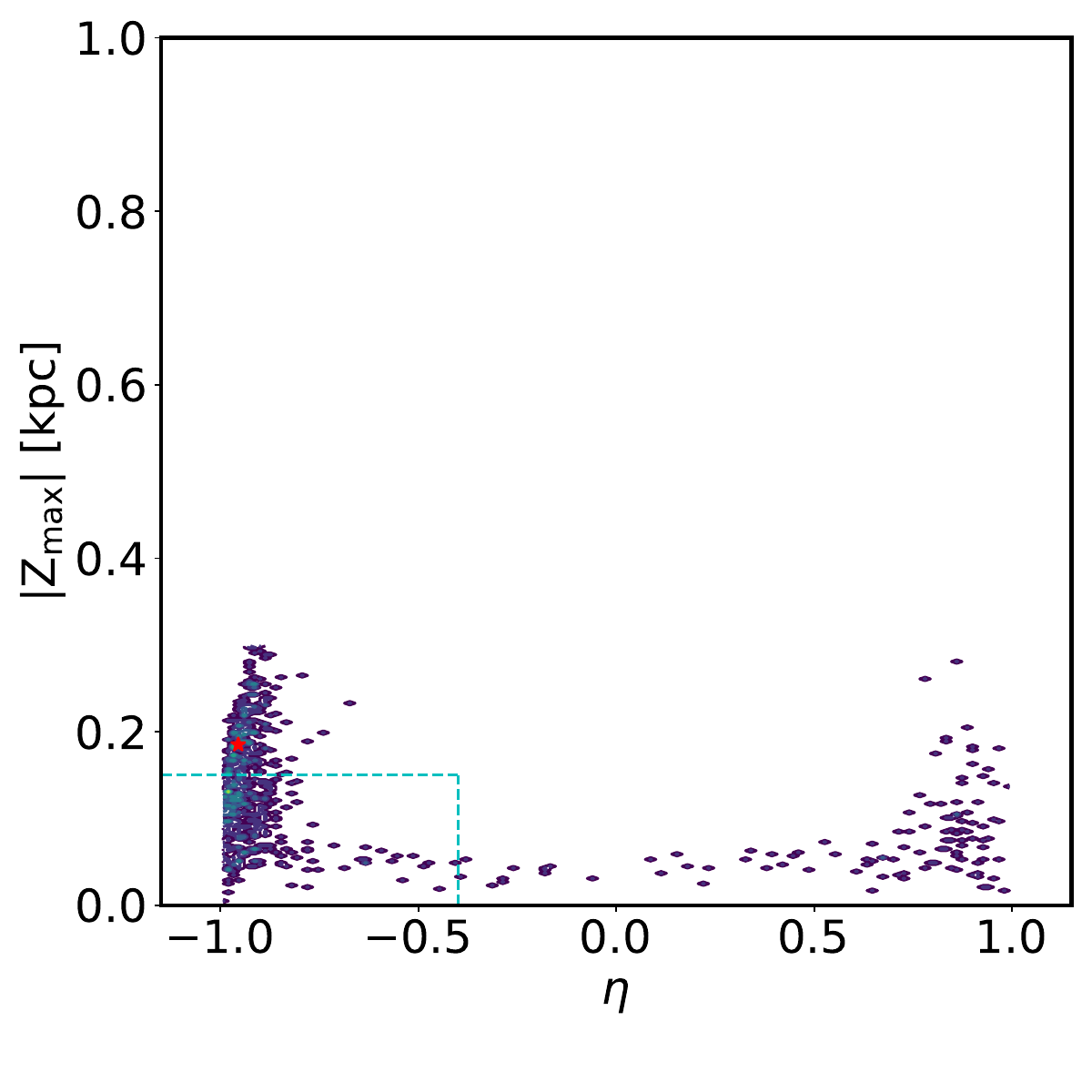}
    \caption{Same as figure ~\ref{GCC-a_ze} but for GCC-c.}\label{app-c_ze}
\end{figure}

For GCC-c, Fig.~\ref{app-c_ze} shows that the Cepheid is extremely prograde but that its orbit reaches heights outside the NSD size, in agreement with our analysis in Sect.~\ref{ceph-c}.

\section{Multi band light curves}\label{multilc}

\begin{figure*}
   \centering
     \includegraphics[width=\textwidth]{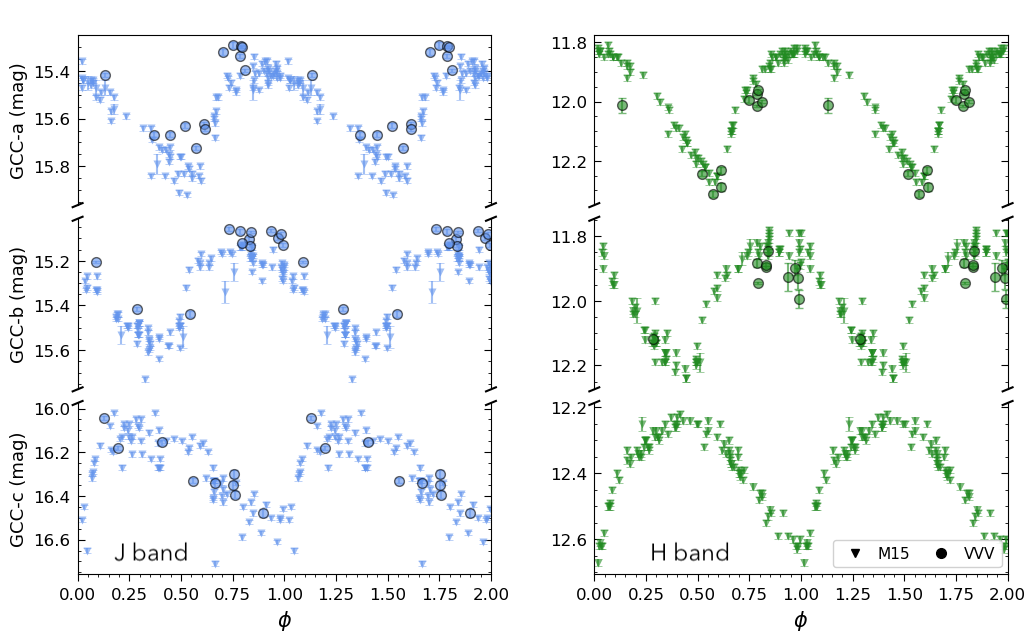}
    \caption{Light curves for, top to bottom, GCC-a, GCC-b and GGC-c, in the J (left) and H band (right). Filled triangles are the data from \citet{matsunaga+15}, while circles are from VVV.}
    \label{fig:LCs_JH}
\end{figure*}

We provide here, in Fig.~\ref{fig:LCs_JH}, the J and H band light curves for the variables GCC-a, GCC-b and GCC-c, from top to bottom in the figure. The literature data from \citet{matsunaga+15} (triangles) were complemented with data from the VVV survey (circles). GCC-c was not recovered in the H band images of VVV. While VVV data add just a small number of points, and they show larger scatter, they do strengthen the original period determination.

\end{appendix}

\end{document}